\documentclass[aps,prx,showpacs,amsmath,amssymb,amsfonts,lengthcheck,twocolumn,longbibliography,floatfix,superscriptaddress]{revtex4-2}
\usepackage[utf8]{inputenc}
\usepackage[top=0.75in,bottom=0.75in,left=0.75in,right=0.75in]{geometry}
\usepackage{graphicx}
\usepackage{subfigure}
\usepackage{amsthm}
\usepackage{soul}

\usepackage{verbatim}
\usepackage{dcolumn}% Align table columns on decimal point
\usepackage{bm}% bold math
\usepackage{epsf}
\usepackage{color}
\usepackage[colorlinks=true,citecolor=blue,linkcolor=blue,urlcolor=blue]{hyperref}%
\usepackage{xcolor}
\usepackage{dsfont}
\newcommand{\bra}[1]{\left\langle #1\right|}
\newcommand{\ket}[1]{\left|#1\right\rangle}
\newcommand{\braket}[2]{\left\langle #1|#2\right\rangle}

\newcommand{\bla}{bla\\bla\\bla\\bla\\bla}

\newcommand{\PRL}{Phys. Rev. Lett. }

\newcommand{\mb}[1]{\mbox{\boldmath$#1$}}

\newcommand{\rev}[1]{{ #1}}

\begin{document}

%\title{Shortcuts to adiabaticity for discrete state Markov systems}
\title{\rev{Shortcuts in stochastic systems and control of biophysical processes}}

\author{Efe Ilker}
\email{ilker@pks.mpg.de}
\affiliation{Laboratoire Physico-Chimie Curie, Institut Curie, PSL Research University, CNRS UMR 168, Paris, France}
\affiliation{Sorbonne Universités, UPMC Univ. Paris 06, Paris, France}
\affiliation{Max Planck Institute for the Physics of Complex Systems, 01187 Dresden, Germany}
\author{\"Ozen\c{c} G\"ung\"or}
\affiliation{ISO/CERCA/Department of Physics, Case Western Reserve University, Cleveland, OH, 44106, USA}
\author{Benjamin Kuznets-Speck}
\affiliation{Department of Physics, Case Western Reserve University, Cleveland, OH, 44106, USA}
\affiliation{Biophysics Graduate Group, University of California, Berkeley, CA 94720, USA}
\author{Joshua Chiel}
\affiliation{Department of Physics, Case Western Reserve University, Cleveland, OH, 44106, USA}
\affiliation{Department of Physics, University of Maryland, College Park, Maryland 20742, USA}
\author{Sebastian Deffner}
\email{deffner@umbc.edu}
\affiliation{Department of Physics, University of Maryland, Baltimore County, Baltimore, MD 21250, USA}
\affiliation{Instituto de F\'isica `Gleb Wataghin', Universidade Estadual de Campinas, 13083-859, Campinas, S\~{a}o Paulo, Brazil}
\author{Michael Hinczewski}
\email{michael.hinczewski@case.edu}
\affiliation{Department of Physics, Case Western Reserve University, Cleveland, OH, 44106, USA}

%\date{\empty}

\begin{abstract}
The biochemical reaction networks that regulate living systems are all stochastic to varying degrees.  The resulting randomness affects biological outcomes at multiple scales, from the functional states of single proteins in a cell to the evolutionary trajectory of whole populations. Controlling how the distribution of these outcomes changes over time---via external interventions like time-varying concentrations of chemical species---is a complex challenge. In this work, we show how counterdiabatic (CD) driving, first developed to control quantum systems, provides a versatile tool for steering biological processes.  We develop a practical graph-theoretic framework for CD driving in discrete-state continuous-time Markov networks.  \rev{Though CD driving is limited to target trajectories that are instantaneous stationary states, we show how to generalize the approach to allow for non-stationary targets and local control---where only a subset of system states are targeted.  The latter is particularly useful for biological implementations where there may be only a small number of available external control knobs, insufficient for global control.  We derive simple graphical criteria for when local versus global control is possible.  Finally, we illustrate the formalism with global control of a genetic regulatory switch and local control in chaperone-assisted protein folding.  The derived control protocols in the chaperone system closely resemble natural control strategies seen in experimental measurements of heat shock response in yeast and {\it E. coli}.}
\end{abstract}

\maketitle

%\section{Introduction}

A fundamental dichotomy for biological processes is that they are both intrinsically stochastic and tightly controlled.  The stochasticity arises from the random nature of the underlying biochemical reactions, and has significant consequences in a variety of contexts: gene expression~\cite{paulsson2005models}, motor proteins~\cite{mugnai2020theoretical}, protein folding~\cite{beauchamp2012simple}, all the way up to the ecological interactions and evolution of entire populations of organisms~\cite{sella2005application,nichol2015steering}.  Theories for such systems often employ discrete state Markov models (or continuum analogues like Fokker-Planck equations) which describe how the probability distribution of system states evolves over time.  On the other hand, biology utilizes a wide array of control knobs to regulate such distributions, most often through time-dependent changes in the concentration of chemical species that influence state transition rates.  In many cases these changes occur due to environmental cues---either threatening or beneficial---and the system response must be sufficiently fast to avoid danger or gain advantage.

The interplay of randomness and regulation naturally leads us to ask about the limits of control:  to what extent can a biological system be driven through a prescribed trajectory of probability distributions over a finite time interval?  Beyond curiosity over whether nature actually tests these limits {\it in vivo}, this question also arises in \rev{other contexts.  In synthetic biology~\cite{khalil2010synthetic} one may want to precisely specify the probabilistic behavior of genetic switches or other regulatory circuit components in response to a stimulus.

Control of a system is generally easiest to describe and quantify if the perturbation is applied slowly (adiabatically).  The advantage of this assumption is that, at each moment of the control protocol, the approximate form of the state probability distribution is known from equilibrium thermodynamics.  However} in natural settings, responses to rapid environmental changes may entail sharp changes in the concentrations of biochemical components. For instance, an ambient temperature increase of even a few degrees can significantly increase the probability that proteins misfold and aggregate. In response to such ``heat shock'', cells quickly upregulate the number of chaperones---specialized proteins that facilitate unfolding or disaggregating misfolded
proteins~\cite{lorimer1996,thirumalai2001,kerner2005,santra2017,richter2010,roncarati2017}.

\begin{figure}[t]
\includegraphics[width=\columnwidth]{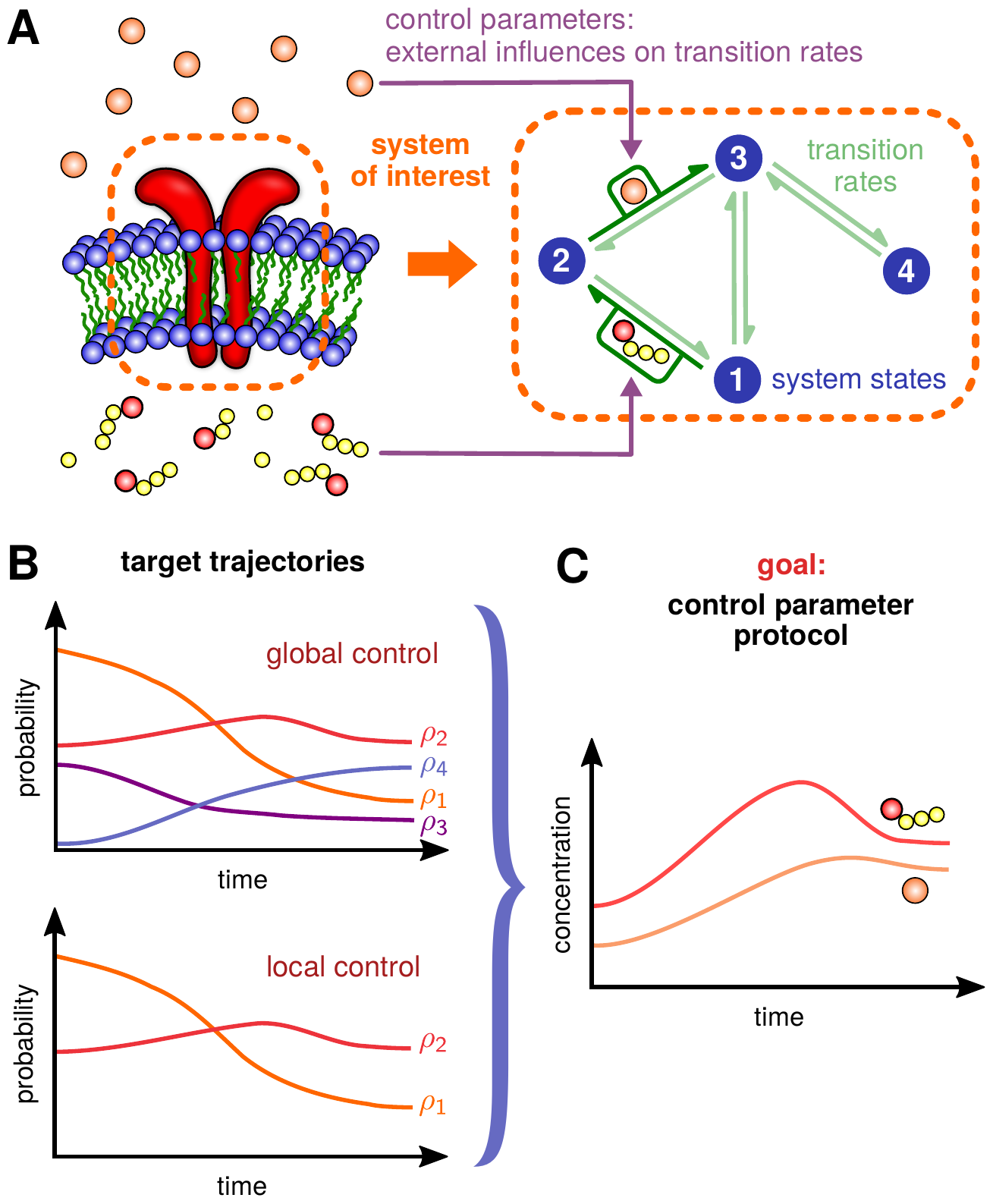}
\caption{\rev{Schematic of the biological control problem:  A) A system of interest (here a membrane receptor protein) is described via a network of transition rates between discrete states.  Certain rates may be influenced by factors external to the system, which we denote as control parameters.  For biochemical systems these are often concentrations of chemical species (ligands, ATP, etc.) or environmental factors like temperature.  B) We consider two types of control:  {\it global control}, where we require the probability of every state to follow a target trajectory over a finite time interval; and {\it local control}, where we impose this requirement on only a subset of states.  C) In either case, the goal is to find whether control is possible for a given target, and if so calculate the control parameter protocol that forces the system to follow the target trajectory.}}\label{biocartoon}
\end{figure}

There is no guarantee that the quasi-equilibrium assumption holds throughout such a process, and thus the standard tools of equilibrium or near-equilibrium thermodynamics (i.e. linear response theory) do not necessarily apply.  \rev{If the system is  driven over a finite-time interval, subject to fluctuations that take it far from equilibrium, can there still be a degree of control?  We can pose this question more concretely, as illustrated in Fig.~\ref{biocartoon}A.  A biological system is typically part of a larger complex of interacting components.  If we focus on a system of interest, and describe it via a discrete state Markov model, the transition rates between states may depend on factors external to the system, like concentrations of ligands that bind to the system, or energy molecules like ATP that are required to fuel certain reactions.  In certain experimental or synthetic biology contexts these factors may be under direct human control, but in natural contexts they are often the product of autonomous processes outside the system of interest, like the temperature fluctuations that lead to heat shock.  In either case we will denote these external factors for simplicity as control parameters, and investigate their influence on the system dynamics.  We will consider a specific question of controllability:  can one find a control protocol (a time-dependent function of the parameters) such that the probability distribution of system states follows a certain sequence of target distributions over a finite time interval?  We can define two types of control (Fig.~\ref{biocartoon}B):  {\it global control}, where we demand the probability of every state in the system follow a chosen trajectory; and {\it local control}, where we only care about a subset of states following the target, and allow the remainder to have arbitrary dynamics.  Ideally, we would like criteria for what kinds of target trajectories are achievable in a given system, and a procedure to calculate the protocol if the target is possible (Fig.~\ref{biocartoon}C).  Answering these questions would not only give us new tools to precisely manipulate biological systems in experiments, but shed light on dynamical constraints {\it in vivo}.  For example, by exploring how controllability depends on the duration of the target trajectory, one can investigate limits on how quickly a system can alter its state distribution in response to an external environmental change.

Interestingly, for one particular class of target trajectories---forcing the system to mimic quasi-equilibrium behavior---}the situation strongly resembles questions from quantum
control and quantum thermodynamics \cite{Deffner2019book}, where a new
line of research has been dubbed ``shortcuts to adiabaticity''. In
recent years a great deal of theoretical and experimental work has
been dedicated to mathematical tools and practical schemes to suppress
nonequilibrium excitations in finite-time, nonequilibrium
processes. To this end, a variety of techniques have been developed:
 the use of dynamical invariants
\cite{Chen2010}, the inversion of scaling laws \cite{Campo2012}, the
fast-forward technique
\cite{Masuda2010,Masuda2011,Torrontegui2012,Torrontegui2012a,Masuda2014a,Kiely2015,Deffner2015NJP,Jarzynski2017PRE},
optimal protocols from optimal control theory
\cite{Chen2011a,Stefanatos2013,Campbell2014,Deffner2014JPB}, optimal
driving from properties of quantum work statistics
\cite{Xiao2014}, ``environment'' assisted methods \cite{Masuda2014,Touil2021},
using the properties of Lie algebras \cite{Torrontegui2014}, and
approximate methods such as linear response theory
\cite{Bonanca2014JCP,Acconcia2015,Bonanca2018PRE,Deffner2020EPL}, fast quasistatic
dynamics \cite{Garaot2015}, or time-rescaling \cite{Bernardo2020,Roychowdhury2021}, to name just a few. See
Refs.~\cite{Torrontegui2013,guery2019sta} and references therein for
comprehensive reviews of these techniques.

Among this plethora of different approaches, \emph{counterdiabatic} (CD) or \emph{transitionless
quantum driving} stands out, since it is the only method that
suppresses excitations away from the adiabatic manifold at all
instants.  In this paradigm~\cite{Demirplak2003, Demirplak2005,
  Berry2009,Deffner2014PRX} one considers a time-dependent Hamiltonian
$H_0(t)$ with instantaneous eigenvalues $\{\epsilon_n(t)\}$ and
eigenstates $\{\ket{n(t)}\}$. In the adiabatic limit no transitions
between eigenstates occur~\cite{Messiah1966}, and each eigenstate
acquires only a time-dependent phase that can be separated into a
dynamical and a geometric contribution~\cite{Berry1984}.  In other words, if we start in a particular eigenstate $\ket{n(0)}$ at $t=0$, we remain in the corresponding instantaneous eigenstate $\ket{n(t)}$ at all later times, up to a phase.  The goal of CD driving is to make the system follow the same target trajectory of eigenstates as in the adiabatic case, but over a finite time.

To accomplish this, a CD Hamiltonian $H(t)$ can be constructed, such that
the adiabatic approximation associated with $H_0(t)$ is an exact
solution of the dynamics generated by $H(t)$ under the time-dependent
Schr\" odinger equation. It is reasonably easy to derive that
time-evolution under \cite{Demirplak2003, Demirplak2005, Berry2009},
\begin{equation}
\label{q01}\begin{split}
H(t)&= H_0(t)+H_1(t)\\
&=H_0(t)+ i\hbar\sum_n\left(\ket{\partial_tn}\bra{n}-\braket{n}{\partial_t n}\ket{n}\bra{ n}\right),
\end{split}
\end{equation}
maintains the system on the adiabatic manifold. Note that it is the
{\it auxiliary Hamiltonian} $H_1(t)$ that enforces evolution along the
adiabatic manifold of $H_0(t)$: if a system is prepared in an
eigenstate $\ket{n(0)}$ of $H_0(0)$ and subsequently evolves under
$H(t)$, then the term $H_1(t)$ effectively suppresses the
non-adiabatic transitions out of $\ket{n(t)}$ that would arise in the
absence of this term.

To date, a few dozen experiments have implemented and utilized such
shortcuts to adiabaticity to, for instance, transport ions or load
BECs into an optical trap without creating parasitic excitations
\cite{guery2019sta}. However, due to the mathematical complexity of
the auxiliary Hamiltonian \eqref{q01}, counterdiabatic driving has
been restricted to ``simple'' quantum systems. Note that in order to compute
$H_1(t)$ one requires the instantaneous eigenstates of the unperturbed
Hamiltonian, which is practically, conceptually, and numerically a
rather involved task.

On the other hand, the scope of CD driving is not limited to the quantum realm.  Because of the close mathematical analogies between classical stochastic systems and quantum mechanics, it was recently recognized that the CD paradigm can
also be formalized for classical scenarios \cite{Jarzynski2013,Deffner2014PRX,Suri_Chris_escort1,Martinez2016,patra2017,Jarzynski2017PRE,Patra2017JPC, swift_Frim,plata2021taming}.  The classical analogue of driving a system along a target trajectory of eigenstates is a trajectory of instantaneous stationary distributions.  Last year, our group and collaborators developed the first biological application of CD driving:  controlling the distribution of genotypes in an evolving cellular population via external drug protocols~\cite{iram2021controlling}.  This type of ``evolutionary steering'' has various potential applications, most notably in designing strategies to combat drug resistance in bacterial diseases and tumors.  The CD formalism in this case was built around a multi-dimensional Fokker-Planck model, generalizing the one-dimensional Fokker-Planck approach of Ref.~\cite{patra2017}.

\rev{Our current work generalizes these initial results in two significant ways:  i) We provide a universal framework capable of handling the wide diversity of stochastic models used in biology, taking advantage of graph theory to construct general algorithms that can be applied to discrete state systems of arbitrary complexity.  The discrete state formalism presented here includes the continuum Fokker-Planck theory as a special case.  ii) Our earlier results were limited to target trajectories that were instantaneous equilibrium distributions (CD driving) defined for all states (global control).  Here we relax both those assumptions:  we allow arbitrary target distributions, and the possibility for defining targets on only a subset of states (local control).  Thus our new formalism includes for example the case of fast-forward driving~\cite{plata2021taming}, where the target trajectory begins and ends in equilibrium, but can be arbitrary in between. The usefulness of our method is of course not confined to biology, but is relevant to other classical systems described by Markovian transitions between states.} However biology provides a singularly fascinating context in which to explore driving, both because it sheds light on the possibility of control in complex stochastic systems with many interacting components, and provides an accessible platform for future experimental tests of these ideas.

\paragraph*{Outline:} In Sec.~\ref{thy} we \rev{start with the most basic version of the theory, formulating CD driving for any discrete state Markov model.}  By looking at the properties of the probability current graph associated with the master equation of the model, we can express CD solutions in terms of spanning trees and fundamental cycles of the graph.  Beyond its practical utility, the graphical approach highlights the degeneracy of CD driving:  the potential existence of many distinct, physically realizable CD protocols that drive a system through the same target trajectory of probability distributions.  The graphical approach is schematically summarized in Fig.~\ref{overview}, highlighting the components in the most general form for CD solutions, Eq.~\eqref{gs10}.

\rev{In Sec.~\ref{gen} we show how the formalism can be generalized to arbitrary (non-CD) target trajectories and local control.  This discussion allows us to derive simple graphical criteria for when global versus local control is possible.  The criteria can help us determine what types of target trajectories are achievable in individual biological systems, based solely on the structure of the corresponding Markovian networks.}

In Sec.~\ref{biol} we apply our formalism to two biological examples, \rev{illustrating global and local control respectively.  The first is a repressor-corepressor genetic regulatory switch, and the second a chaperone protein that catalyzes the unfolding of a misfolded protein in response to a heat shock.  The switch provides perhaps the simplest example where the parameters have been experimentally characterized and various driving solutions can be directly tested {\it in vitro}.  For the chaperone system, we highlight the qualitative similarities between local control protocols for rapidly suppressing misfolded proteins and experimental measurements of heat shock response in yeast and {\it E. coli}.}  

Sec.~\ref{con} concludes with connections to other areas of nonequilibrium thermodynamics and questions for future work.

\section{General theory of counterdiabatic driving in discrete state Markov models}\label{thy}

\subsection{Setting up the counterdiabatic driving problem}

\subsubsection{Master equation and the CD transition matrix}

% \subsubsection{Master equation and the counterdiabatic transition matrix}

Consider an $N$-state Markov system described by a vector $\mb{p}(t)$
whose component $p_i(t)$, $i=1,\ldots, N$, is the probability of being
in state $i$ at time $t$.  The distribution $\mb{p}(t)$ evolves under
the master equation~\cite{vankampen,esposito2010},
\begin{equation}\label{1}
\partial_t{\mb{p}}(t) = \Omega(\lambda_t) \mb{p}(t).
\end{equation}
The off-diagonal element $\Omega_{ij}(\lambda_t)$, $i \ne j$, of
the $N \times N$ matrix $\Omega(\lambda_t)$ represents the conditional
probability per unit time to transition to state $i$, given that the
system is currently in state $j$. The diagonal elements
$\Omega_{ii}(\lambda_t) = - \sum_{j \ne i} \Omega_{ji}(\lambda_t)$
ensure each column of the matrix sums to zero~\cite{vankampen}. The
transition rates $\Omega_{ij}(\lambda_t)$ depend on a control protocol: a set of time-varying
external parameters, denoted
collectively by $\lambda(t) \equiv  \lambda_t$. $\Omega(t)$ plays the role of the Hamiltonian $H_0(t)$ in the classical analogy.

The instantaneous stationary probability $\bm{\rho}(\lambda_t)$ associated with $\Omega(\lambda_t)$ is the right eigenvector with eigenvalue zero,
\begin{equation}\label{2}
\Omega(\lambda_t) \bm{\rho}(\lambda_t) = 0.
\end{equation}
When $\lambda_t$ has a non-constant time dependence, $\bm{\rho}(\lambda_t)$ in general is
not a solution to Eq.~\eqref{1}, except in the adiabatic limit when
the control parameters are varied infinitesimally slowly, $\partial_t
\lambda_t \to 0$.  The sequence of distributions $\bm{\rho}(\lambda_t)$ as a function of $\lambda_t$ defines a target trajectory for the system, analogous to the eigenstate trajectory $\ket{n(t)}$ in the quantum version of CD.

Given an instantaneous probability trajectory $\bm{\rho}(\lambda_t)$
defined by Eq.~\eqref{2}, we would like to find a counterdiabatic (CD)
transition matrix $\widetilde{\Omega}(\lambda_t,\dot\lambda_t)$ such
that the new master equation,
\begin{equation}\label{3}
\partial_t{\bm{\rho}}(\lambda_t) = \widetilde{\Omega}(\lambda_t,\dot\lambda_t) \bm{\rho}(\lambda_t),
\end{equation}
evolves in time with state probabilities described by
$\bm{\rho}(\lambda_t)$.  Here $\dot{\lambda}_t \equiv \partial_t
\lambda_t$. We are thus forcing the system to mimic adiabatic time
evolution, even when $\dot{\lambda}_t$ is nonzero. As we will see
below, $\widetilde{\Omega}(\lambda_t,\dot\lambda_t)$ will in general
depend both on the instantaneous values of the control parameters
$\lambda_t$ and their rate of change $\dot{\lambda}_t$.  In the limit
of adiabatic driving we should recover the original transition matrix,
$\widetilde{\Omega}(\lambda_t,\dot\lambda_t \to 0) =
\Omega(\lambda_t)$.  Solving for the CD protocol corresponds to
determining the elements of the
$\widetilde{\Omega}(\lambda_t,\dot\lambda_t)$ matrix in Eq.~\eqref{3}
given a certain $\bm{\rho}(\lambda_t)$.  This corresponds to finding the CD Hamiltonian $H(t)$ of Eq.~\eqref{q01} in the quantum case.

We can look at the counterdiabatic problem as a special case of a more
general question: given a certain time-dependent probability
distribution that is our target, what is the transition matrix of the
master equation for which this distribution is a solution?  In effect,
this is the inverse of the typical approach for the master equation,
where we know the transition matrix and solve for the distribution.

\subsubsection{Representing the system via an oriented current graph}

To facilitate finding CD solutions, we start by expressing the
original master equation of Eq.~\eqref{1} equivalently in terms of
probability currents between states,
\begin{equation}\label{ff1}
\partial_t p_i(t) = \sum_{j} J_{ij}(t), \quad i=1,\ldots,N
\end{equation}
where the current from state $j$ to $i$ is given by:
\begin{equation}\label{ff1b}
J_{ij}(t) \equiv \Omega_{ij}(\lambda_t) p_{j}(t) - \Omega_{ji}(\lambda_t) p_{i}(t).
\end{equation}
We can interpret any pair of states $(i,j)$ where either
$\Omega_{ij}(\lambda_t)\ne 0$ or $\Omega_{ji}(\lambda_t)\ne 0$ at some
point during the protocol as being connected via an edge on a graph
whose vertices are the states $i = 1,\ldots,N$. \rev{Let $E$ be the number
of edges in the resulting graph.  Define a numbering $\alpha =
1,\ldots,E$ and an arbitrary orientation for the edges such that each
$\alpha$ corresponds to a specific edge and choice of current
direction.}  For example, if edge $\alpha$ was between states $(i,j)$,
and the choice of direction was from $j$ to $i$, then we can define
current $J_\alpha(t) \equiv J_{ij}(t)$ for that edge.
Alternatively if the choice of direction was from $i$ to $j$, then
$J_\alpha(t) \equiv J_{ji}(t) = -J_{ij}(t)$.  \rev{We denote rates $\Omega_{ij}(\lambda_t)$ oriented
parallel to the edge direction as forward rates, and those oriented opposite as backward rates.}
In this way we associate the master equation with a directed graph,
a simple example of which is illustrated in Fig.~\ref{graphs}.
Eq.~\eqref{ff1} can be rewritten in terms of the oriented currents
$J_\alpha(t)$ as
\begin{equation}\label{f2}
\partial_t {\bm{p}}(t) = \nabla \bm{J}(t),
\end{equation}
where $\bm{J}(t)$ is an $E$-dimensional vector with components
$J_\alpha(t)$, and $\nabla$ is an $N \times E$ dimensional
matrix known as the incidence matrix of the directed
graph~\cite{deo2017graph} (closely related to the stoichiometric
matrix of Ref.~\cite{rao2016nonequilibrium}).  \rev{$\nabla$ is given by
\begin{equation}\label{f2b}
\nabla = \nabla^- - \nabla^+,
\end{equation}
where the components of the two matrices $\nabla^\pm$ are defined as:
\begin{equation}\label{f3}
\begin{split}
\nabla^-_{i\alpha} &= 
\begin{cases}
1 & \text{if the direction of edge $\alpha$ is toward $i$}\\
0 & \text{otherwise}
\end{cases},\\
\nabla^+_{i\alpha} &= 
\begin{cases}
1 & \text{if the direction of edge $\alpha$ is away from $i$}\\
0 & \text{otherwise}
\end{cases}.\\
\end{split}
\end{equation}
The $\alpha$th column of $\nabla$ contains a single $1$ and
a single $-1$, since each edge must have an origin and a destination
state.  With these definitions, Eq.~\eqref{ff1b} can be recast as a relation between the vectors $\bm{J}(t)$ and $\bm{p}(t)$,
\begin{equation}
    \label{m1ex}
    \bm{J}(t) = G(\lambda_t) \bm{p}(t),
\end{equation}
where the $E \times N$ dimensional matrix $G(\lambda_t)$ is given by
\begin{equation}
    \label{m2ex}
    G(\lambda_t) = \text{diag}(\bm{k}^+(\lambda_t)) {\nabla^+}^T - \text{diag}(\bm{k}^-(\lambda_t)) {\nabla^-}^T.
\end{equation}
Here $\text{diag}(\bm{k}^+(\lambda_t))$ is an $E \times E$ dimensional diagonal matrix where the diagonal is $\bm{k}^+(\lambda_t)$, the vector of forward rates associated with each edge.  For example $k^+_{\alpha}(\lambda_t) = \Omega_{ij}(\lambda_t)$ if the $\alpha$th edge is oriented from $j$ to $i$.  Similarly $\bm{k}^-(t)$ is the vector of backward rates.  Comparing Eq.~\eqref{1} to Eqs.~\eqref{f2}-\eqref{m2ex}, we see that the matrix $\Omega(\lambda_t) = \nabla G(\lambda_t)$.  In the special case where the matrix $G(\lambda_t)$ has a right singular vector with singular value zero, we say that the rates in the system satisfy instantaneous detailed balance.  We will refer to $\bm{k}^\pm(\lambda_t)$ as the ``original'' rate protocol for the system, since they determine the original transition matrix $\Omega(\lambda_t)$ and hence the target $\bm{\rho}(\lambda_t)$ via Eq.~\eqref{2}.  Throughout the text we will use ``original'' to consistently describe quantities associated with $\Omega(\lambda_t)$.  On the other hand quantities associated with the CD matrix $\widetilde{\Omega}(\lambda_t,\dot\lambda_t)$, like the CD rates $\bm{\widetilde{k}}^\pm(\lambda_t)$ described below, will always have tildes to distinguish them from the original case.
}

Conservation of probability is enforced by summing over rows in Eq.~\eqref{f2}, since $\sum_i \nabla_{i\alpha} = 0$, and so
$\sum_{i=1}^N \partial_t p_i(t) = 0$.  Since any given row of
Eq.~\eqref{f2} is thus linearly dependent on the other rows, it is
convenient to work in the reduced representation of the equation where
we leave out the row corresponding to a certain reference state (taken
to be state $N$),
\begin{equation}\label{f4}
\partial_t{\widehat{\bm{p}}}(t) = \widehat{\nabla} \bm{J}(t).
\end{equation}
Here $\widehat{\bm{p}}(t) = (p_1(t),\ldots,p_{N-1}(t))$ and the $(N-1)
\times E$ dimensional reduced incidence matrix $\widehat{\nabla}$ is
equal to $\nabla$ with the $N$th row removed.  Our focus will be on
systems where there is a unique instantaneous stationary probability
vector $\bm{\rho}(\lambda_t)$ at every $t$.  In this case the master
equation necessarily corresponds to a connected graph in the oriented
current picture~\cite{vankampen}.  By a well known result in graph
theory, both the full and reduced incident matrices $\nabla$ and
$\widehat\nabla$ of a connected, directed graph with $N$ vertices have
rank $N-1$~\cite{deo2017graph}.  This means that all $N-1$ rows of
$\widehat\nabla$ are linearly independent for the systems we consider.

Having described the original master equation of Eq.~\eqref{1} in
terms of oriented currents, we can do the same for Eqs.~\eqref{2} and
\eqref{3}.  Let us define the oriented stationary current
$\mathcal{J}_\alpha(t)$ for the distribution
$\bm{\rho}(\lambda_t)$ as follows: if the $\alpha$th edge is oriented
from $j$ to $i$ then
\begin{equation}\label{f5}
\mathcal{J}_\alpha(t) \equiv \Omega_{ij}(\lambda_t) \rho_j(\lambda_t) - \Omega_{ji}(\lambda_t) \rho_i(\lambda_t).
\end{equation}
\rev{In vector form, analogous to Eq.~\eqref{m1ex}, the current is given by
\begin{equation}\label{m3ex}
\bm{\mathcal{J}}(t) = G(\lambda_t) \bm{\rho}(t).
\end{equation}}
The reduced representation of Eq.~\eqref{2} corresponds to
\begin{equation}\label{f6}
\widehat\nabla \bm{\mathcal{J}}(t) = 0.
\end{equation}
\rev{If the system rates satisfy instantaneous detailed balance, $\bm{\mathcal{J}}(t)=0$, since $\bm\rho(t)$ is the right singular vector of $G(\lambda_t)$ with singular value zero.  However our CD approach works for the more general case where $\bm{\mathcal{J}}(t)$ can be nonzero but Eq.~\eqref{f6} is satisfied.  In fact, as we will see in Sec.~II.A, we can also generalize our theory to completely arbitrary (non-CD) target trajectories $\bm{\rho}(t)$ where Eq.~\eqref{f6} no longer holds.}

For the CD master equation, Eq.~\eqref{3}, we define the oriented current
\begin{equation}\label{f7}
\mathcal{\widetilde{J}}_\alpha(t) \equiv \widetilde\Omega_{ij}(\lambda_t,\dot\lambda_t) \rho_j(\lambda_t) - \widetilde\Omega_{ji}(\lambda_t,\dot\lambda_t) \rho_i(\lambda_t).
\end{equation}
The time dependence of ${\mathcal{\widetilde{J}}}_\alpha$ is explicitly
through $\lambda_t$ and $\dot\lambda_t$, but we write it in more
compact form as ${\mathcal{\widetilde{J}}}_\alpha(t)$ to avoid cumbersome notation.  \rev{The analogue of Eq.~\eqref{m1ex} is
\begin{equation}
    \label{m4ex}
    \bm{\mathcal{\widetilde{J}}}(t) = \widetilde{G}(t) \bm{\rho}(t),
\end{equation}
where $\widetilde{G}(t)$ has the same structure as Eq.~\eqref{m2ex} but with forward/backward rate vectors $\bm{\widetilde{k}}^\pm(t)$ corresponding to the CD rates $\widetilde\Omega_{ij}(\lambda_t,\dot\lambda_t)$.}  Finally, Eq.~\eqref{3} can be expressed as
\begin{equation}\label{f8}
    \partial_t \widehat{\bm{\rho}}(\lambda_t) = \widehat\nabla \bm{\mathcal{\widetilde{J}}}(t).
\end{equation}

\begin{figure*}
    \centering
    \includegraphics[width=0.95\textwidth]{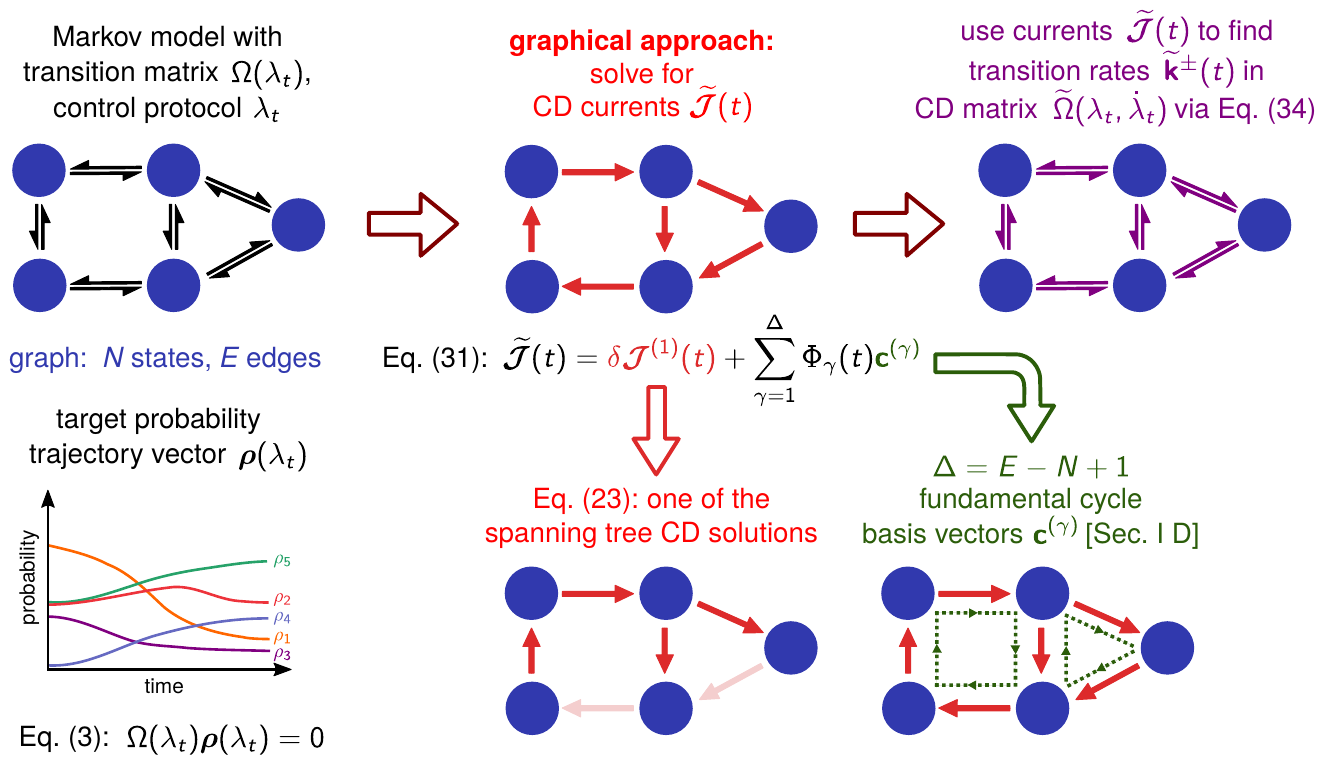}
    \caption{Overview of the graphical approach for deriving CD solutions.  We start with a Markov model defined by a transition matrix $\Omega(\lambda_t)$ dependent on the control protocol $\lambda_t$.  Associated with this is a graph with $N$ states, $E$ edges, and a target trajectory $\bm{\rho}(\lambda_t)$ consisting of instantaneous stationary states of $\Omega(\lambda_t)$.  The eventual goal is to find the CD transition matrix $\widetilde{\Omega}(\lambda_t,\dot\lambda_t)$ where $\bm{\rho}(\lambda_t)$ is the solution to the associated master equation, Eq.~\eqref{3}.  To facilitate this, we must first find the CD currents ${\bm{\mathcal{\widetilde J}}}(t)$, the main goal of the graphical approach.  The most general form of the solution for ${\bm{\mathcal{\widetilde J}}}(t)$ is given by Eq.~\eqref{gs10}, and consists of two components: (i) a spanning tree CD solution $\delta {\bm{\mathcal{J}}}^{(1)}(t)$, given by Eq.~\eqref{gs3} and derived via the procedure outlined in Sec.~\ref{sec:graphsol}; (ii) a linear combination of the fundamental basis cycle vectors $\bm{c}^{(\gamma)}$, $\gamma = 1,\ldots,\Delta$, where $\Delta = E-N+1$, as described in Sec.~\ref{sec:cycbasis}.  The coefficient functions $\Phi_\gamma(t)$ are arbitrary.  \rev{Once the CD currents ${\bm{\mathcal{\widetilde J}}}(t)$ are known, we can use Eq.~\eqref{sol3} to solve for the CD transition rates $\bm{\widetilde{k}}^\pm(t)$ that determine $\widetilde{\Omega}(\lambda_t,\dot\lambda_t)$.}}
    \label{overview}
\end{figure*}

\subsubsection{Counterdiabatic current equation}

Subtracting Eq.~\eqref{f6} from Eq.~\eqref{f8} we find
\begin{equation}\label{f9}
\partial_t \widehat{\bm{\rho}}(\lambda_t) = \widehat\nabla \delta\bm{\mathcal{J}}(t),
\end{equation}
where $\delta\bm{\mathcal{J}}(t) \equiv \bm{\mathcal{\widetilde J}}(t)
- \bm{\mathcal{J}}(\lambda_t)$ is the difference between the CD and
stationary current vectors.  For the CD problem, we are given the original matrix
elements $\Omega_{ij}(\lambda_t)$ and thus also have the
corresponding stationary distribution values $\rho_i(\lambda_t)$ and
stationary currents $\mathcal{J}_\alpha(\lambda_t)$.  What we need to
determine, via Eq.~\eqref{f9}, are the CD currents
$\bm{\mathcal{\widetilde J}}(t)$.  \rev{The following Secs. I.B through I.D detail the procedure for finding these currents. Once we know $\bm{\mathcal{\widetilde J}}(t)$, Sec.~\ref{solve} shows how to use Eq.~\eqref{m4ex} to solve for $\bm{\widetilde{k}}^\pm(t)$, or equivalently the CD matrix transition rates $\widetilde\Omega_{ij}(\lambda_t,\dot\lambda_t)$.}  By construction,
these satisfy Eq.~\eqref{3}, and hence define a CD protocol for the
system.

As a first step, let us consider the invertibility of Eq.~\eqref{f9}
to solve for $\delta\bm{\mathcal{J}}(t)$.  The
$(N-1) \times E$ dimensional matrix $\widehat\nabla$ is generally
non-square: $N(N-1)/2 \ge E \ge N-1$ for a connected graph.  Only in
the special case of tree-like graphs (no loops) do we have $E = N-1$
and a square $(N-1) \times (N-1)$ matrix $\widehat\nabla$.  Since the
rank of $\widehat\nabla$ is $N-1$, as mentioned above, for tree-like
graphs $\widehat\nabla$ is invertible and Eq.~\eqref{f9} can be solved
without any additional complications:
\begin{equation}\label{f10}
\delta\bm{\mathcal{J}}(t) = \widehat\nabla^{-1} \partial_t \widehat{\bm{\rho}}(\lambda_t) \quad \text{iff\:} E=N-1.
\end{equation}
As described in the next section, the elements of
$\widehat\nabla^{-1}$ for a tree-like graph can be obtained directly
through a graphical procedure, without the need to do any explicit
matrix inversion.

In the case where $E > N-1$, the solution procedure is more involved,
but the end result has a relatively straightforward form: the most
general solution $\delta\bm{\mathcal{J}}(t)$ can
always be expressed as a finite linear combination of a basis of CD
solutions.  How to obtain this basis, and its close relationship to the
spanning trees and fundamental cycles of the graph, is the topic we
turn to next.

\subsection{General graphical solution for the counterdiabatic protocol}\label{sec:graphsol}

\begin{figure}
    \centering
    \includegraphics[width=\columnwidth]{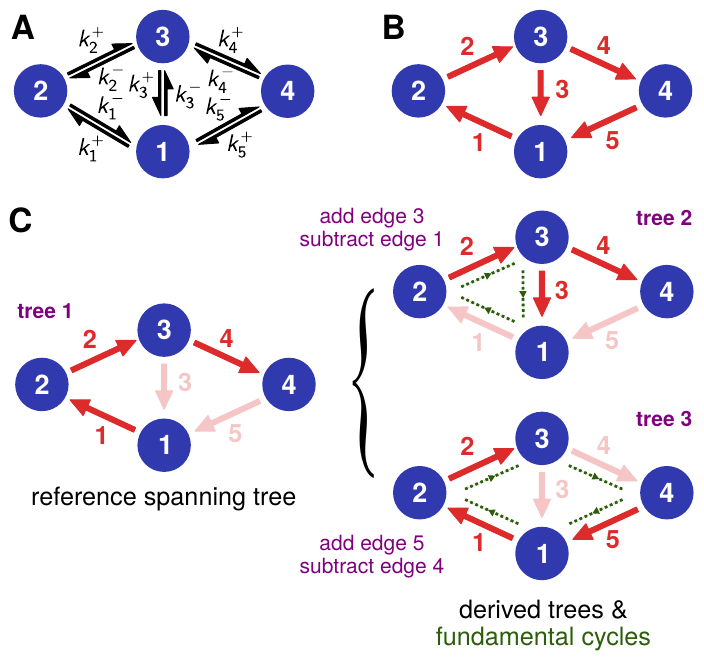}
    \caption{A two-loop discrete state Markov model, with $N=4$ states
      and $E=5$ edges.  A) The black arrows correspond to entries in
      the transition matrix $\Omega(\lambda_t)$: transition rates
      \rev{$k^\pm_i(\lambda_t)$} that depend
      on an external protocol $\lambda_t$.  B) The red arrows labeled \rev{$\alpha$ correspond to the} oriented stationary currents
      $\mathcal{J}_\alpha(\lambda_t)$, defined in Eq.~\eqref{f5}.  C)
      On the left, one of the spanning trees of the graph, chosen to
      be a reference for constructing the tree basis.  Edges deleted to form the tree are shown in faint red.  On the right,
      two trees in this set derived from the reference one.  Each such
      derived tree has a one-to-one correspondence with a fundamental
      cycle of the graph (highlighted in green).}
    \label{graphs}
\end{figure}

The graphical procedure described in this and the following two sections, culminating in the general solution of Eq.~\eqref{gs10}, is summarized in Fig.~\ref{overview}.  To illustrate the procedure concretely, we will use the
two-loop system shown in Fig.~\ref{graphs}A as an example, where
$N=4$, $E=5$.  The solution for this case is relevant to the
biophysical model for chaperone-assisted protein folding discussed
i\rev{n Sec. III.B}.  Fig.~\ref{graphs}A shows the rates
\rev{$k^\pm_\alpha(\lambda_t)$} that determine the transition
matrix $\Omega(\lambda_t)$, and Fig.~\ref{graphs}B labels the oriented
stationary currents $\mathcal{J}_\alpha(t)$, $\alpha
=1,\ldots,E$.

Every connected graph has a set of spanning trees: subgraphs formed by
removing $\Delta \equiv E - N +1$ edges such that the remaining $N-1$
edges form a tree linking together all the $N$ vertices.  The number
${\cal T}$ of such spanning trees is related to the reduced incidence
matrix through Kirchhoff's matrix tree theorem~\cite{deo2017graph},
${\cal T} = \det\left(\widehat\nabla \widehat\nabla^T \right)$.  For
the current graph of Fig.~\ref{graphs}B, this matrix is
\rev{\begin{equation}\label{gs1}
\widehat\nabla = \begin{pmatrix}
-1 & 0 & 1 & 0 & 1\\
1 & -1 & 0 & 0 & 0\\
0 & 1 & -1 & -1 & 0
\end{pmatrix},
\end{equation}}
and the number of trees is thus ${\cal T} = 8$.

Let us select one spanning tree to label as the reference tree.  The
choice is arbitrary, since any spanning tree can be a valid starting
point for constructing the basis.  The left side of Fig.~\ref{graphs}C
shows one such tree chosen for the two-loop example.  Here $\Delta =
2$, so we have removed two edges: \rev{${\cal J}_3$} and ${\cal J}_5$.  From
this reference tree we can derive $\Delta$ other distinct spanning
trees using the following method: 1) Take one of the $\Delta$ edges
that were removed to get the reference tree, and add it back to the
graph.  2) This creates a loop in the graph, known as a fundamental
cycle (highlighted in green in
Fig.~\ref{graphs}C)~\cite{deo2017graph}. 3) Remove one of the other
edges in that loop (not the one just added), such that the graph
returns to being a spanning tree.  This new tree is distinct from the
reference because it contains one of the $\Delta$ edges not present in
the reference tree.  For example, in the top right of
Fig.~\ref{graphs}C, we added back edge \rev{3}, forming the fundamental
cycle on the left loop.  We then delete edge \rev{1} from this loop,
creating spanning tree 2. A similar procedure is used to construct tree 3.

We denote the $\Delta +1$ trees (one reference + $\Delta$ derived
trees) constructed in this manner as the tree basis.  We will label
the trees in the basis set with $\gamma =1,\ldots,\Delta +1$, where
$\gamma=1$ corresponds to the reference.  In general, this
basis is a subset of all possible trees, since ${\cal T} \ge \Delta
+1$.  To every tree in the basis, we will associate a CD solution as
follows.  Let $\delta\bm{\mathcal{J}}^{(\gamma)}(t)$ be a current difference vector that satisfies
Eq.~\eqref{f9}, but with the constraint that at every edge $\alpha$
that is not present in the $\gamma$th tree, we have
$\delta\mathcal{J}^{(\gamma)}_\alpha(t) = 0$.
We call this a {\it fixed current} constraint, since it corresponds to
not being able to perturb the current associated with that edge via
external control parameters.  For example, imposing the restriction
$\Omega_{ij} = \widetilde\Omega_{ij}$ and $\Omega_{ji} =
\widetilde\Omega_{ji}$ for the pair $(i,j)$ associated with edge
$\alpha$ would make make
$\delta\mathcal{J}^{(\gamma)}_\alpha(t) = 0$.

To find $\delta\bm{\mathcal{J}}^{(\gamma)}(t)$,
consider the $(N-1) \times E$-dimensional reduced incidence matrix
$\widehat{\nabla}$ of the original graph; for example, Eq.~\eqref{gs1}
in the case of the two-loop graph of Fig.~\ref{graphs}B.  For a given
spanning tree $\gamma$, we can construct an $(N-1) \times (N-1)$
submatrix $\widehat{\nabla}^{(\gamma)}$ from $\widehat{\nabla}$ by
choosing the $N-1$ columns in $\widehat{\nabla}$ that correspond to
edges present in $\gamma$.  This submatrix
$\widehat{\nabla}^{(\gamma)}$ is equal to the reduced incidence matrix
of the spanning tree $\gamma$.  Hence we know that it has rank $N-1$
and there exists an inverse $[\widehat{\nabla}^{(\gamma)}]^{-1}$.  Let
us now construct a ``stretched inverse'': an $E \times
(N-1)$-dimensional matrix $[\widehat{\nabla}^{(\gamma)}]^{-1}_S$ where
the rows are populated by the following rule.  If the row corresponds
to one of the $\Delta$ edges that was removed from the original graph
to get the tree $\gamma$, it is filled with zeros; otherwise, it is
filled with the corresponding row of
$[\widehat{\nabla}^{(\gamma)}]^{-1}$.  For the three trees in
Fig.~\ref{graphs}C, labeled $\gamma=1,2,3$ clockwise from left, the
matrices $[\widehat{\nabla}^{(\gamma)}]^{-1}_S$ have the following
form:
\rev{\begin{equation}\label{gs2}
\begin{split}
[\widehat\nabla^{(1)}]^{-1}_S &= \begin{pmatrix}
-1 & 0 & 0\\
-1 & -1 & 0\\
0 & 0 & 0\\
-1 & -1 & -1\\
0 & 0 & 0\\
\end{pmatrix},\\
[\widehat\nabla^{(2)}]^{-1}_S &= \begin{pmatrix}
0 & 0 & 0\\
0 & -1 & 0\\
1 & 0 & 0\\
-1 & -1 & -1\\
0 & 0 & 0\\
\end{pmatrix},
\quad [\widehat\nabla^{(3)}]^{-1}_S = \begin{pmatrix}
0 & 1 & 1\\
0 & 0 & 1\\
0 & 0 & 0\\
0 & 0 & 0\\
1 & 1 & 1\\
\end{pmatrix}.
\end{split}
\end{equation}}
Moreover, it turns out one does not have to explicitly write down or invert
$\widehat\nabla^{(\gamma)}$ in order to find the elements of
$[\widehat\nabla^{(\gamma)}]^{-1}_S$.  We can take advantage of a
known graphical procedure for constructing inverse reduced incidence
matrices of connected tree-like graphs~\cite{resh1963inverse,
  bevis1981integer}.  To determine the $i$th column of the matrix
$[\widehat\nabla^{(\gamma)}]^{-1}_S$, start at the reference state
(the state removed when constructing the reduced incidence matrix
$\widehat\nabla$, which in our case is always state $N$).  Among the
edges of the spanning tree $\gamma$, there is a unique path that
connects state $N$ to state $i$.  Following that path, if you
encounter the current arrow $\mathcal{J}_\alpha$ oriented in the
direction of the path, put a $+1$ in the row of
$[\widehat\nabla^{(\gamma)}]^{-1}_S$ corresponding to
$\mathcal{J}_\alpha$.  Similarly if the current arrow is oriented
opposite to the path, put a $-1$.  All other entries in the $i$th
column (current arrows not on the path, or not in the spanning tree)
are set to zero.  For example, consider the second column of
$[\widehat\nabla^{(1)}]^{-1}_S$ in Eq.~\eqref{gs2}.  This corresponds
to the path from state $N=4$ to state 2 in the tree on the left of
Fig.~\ref{graphs}C.  This includes edges \rev{4 and 2}, with the arrows
along those edges all oriented opposite to the path.  Hence the column
has a $-1$ at the \rev{4th and 2nd} rows, and all other entries are set
to zero.

By construction, each matrix $[\widehat\nabla^{(\gamma)}]^{-1}_S$
acts as a right pseudoinverse of $\widehat{\nabla}$,
satisfying $\widehat{\nabla} [\widehat\nabla^{(\gamma)}]^{-1}_S =
I_{N-1}$, where $I_{N-1}$ is the $(N-1) \times (N-1)$ dimensional
identity matrix.  We can now write down a solution for
$\delta\bm{\mathcal{J}}^{(\gamma)}(t)$,
\begin{equation}\label{gs3}
\delta\bm{\mathcal{J}}^{(\gamma)}(t) = [\widehat\nabla^{(\gamma)}]^{-1}_S \partial_t \widehat{\bm{\rho}}(\lambda_t).
\end{equation}
If we act from the left on both sides by $\widehat{\nabla}$, we see
that this form satisfies Eq.~\eqref{f9}. The $\alpha$th row of
of $[\widehat\nabla^{(\gamma)}]^{-1}_S$ is zero if edge $\alpha$ corresponds to a fixed current constraint (edge not present in the
tree $\gamma$).  Thus $\delta\mathcal{J}^{(\gamma)}_\alpha(t) =0$ for these $\alpha$.  Not only do the
vectors $\delta {\bm{\mathcal{J}}}(t)$
associated with the tree basis constitute $\Delta +1$ solutions to
Eq.~\eqref{f9}, they are also linearly independent from one another.
To see this, note that because of the procedure to construct derived
trees (adding back a distinct edge that was removed in the reference
tree), a tree with $\gamma \ge 2$ will have non-zero entry in $\delta
{\bm{\mathcal{J}}}^{(\gamma)}(t)$ in a position
where every other tree (reference or derived) has a zero because of
constraints.  Hence the $\delta
{\bm{\mathcal{J}}}^{(\gamma)}(t)$ vector for
each derived tree is linearly independent from all the other vectors
in the basis.

We also know that any linear combination of solutions to Eq.~\eqref{f9} can be scaled by an overall normalization factor (to make the coefficients sum to one) so that it is also a solution to Eq.~\eqref{f9}.  Hence the following linear combination of basis vectors is a valid solution:
\begin{equation}\label{gs4}
    \delta {\bm{\mathcal{J}}}(t) =
    \sum_{\gamma=1}^{\Delta +1} w_\gamma(t)
    \delta {\bm{\mathcal{J}}}^{(\gamma)}(t),
\end{equation}
Here $w_\gamma(t)$ are any real-valued functions
where $\sum_{\gamma=1}^{\Delta +1} w_\gamma(t)
=1$ at each $\lambda_t$ and $\dot\lambda_t$.  As we argue in the next
section, the tree basis is complete: any CD solution $\delta
{\bm{\mathcal{J}}}(t)$ can be expressed in the
form of Eq.~\eqref{gs4}.  Note that Eq.~\eqref{f10} is a special case
of Eq.~\eqref{gs4}.  When the original graph is tree-like, $\Delta =0$
and there is only one spanning tree ($\gamma=1$), equivalent to the
original graph.  In this case $[\widehat{\nabla}^{(1)}]^{-1}_S =
\widehat{\nabla}^{-1}$ and the sole coefficient function
$w_1(t)=1$ by normalization.

\subsection{Completeness of the tree basis}

To prove that any CD solution can be expressed as a linear combination
of tree basis solutions $\delta
{\bm{\mathcal{J}}}^{(\gamma)}(t)$, let us first
introduce $\Delta$ vectors of the following form:
\begin{equation}\label{gs5}
\bm{\mathcal{V}}^{(\gamma)}(t) = \delta {\bm{\mathcal{J}}}^{(\gamma)}(t) - \delta {\bm{\mathcal{J}}}^{(1)}(t),
\end{equation}
for $\gamma=2,\ldots,\Delta+1$.  Since both basis vectors on the
right-hand side of Eq.~\eqref{gs5} satisfy Eq.~\eqref{f9}, we know
that
\begin{equation}\label{gs6}
\widehat{\nabla}\bm{\mathcal{V}}^{(\gamma)}(t) = \partial_t \widehat{\bm{\rho}}(\lambda_t) - \partial_t \widehat{\bm{\rho}}(\lambda_t) = 0.
\end{equation}
Hence $\bm{\mathcal{V}}^{(\gamma)}(t)$ is a
vector in the null space of $\widehat{\nabla}$.  Moreover since the
basis vectors $\delta
{\bm{\mathcal{J}}}^{(\gamma)}(t)$ are linearly
independent, the set
$\bm{\mathcal{V}}^{(\gamma)}(t)$ constitutes
$\Delta$ linearly independent null vectors of $\widehat{\nabla}$.  We
can find the dimension of the null space,
$\text{nullity}(\widehat\nabla)$, using the rank-nullity theorem:
$\text{rank}(\widehat\nabla) + \text{nullity}(\widehat\nabla) = E$,
where $E$ is the number of columns in $\widehat\nabla$.  Since
$\text{rank}(\widehat\nabla) = N-1$ for a connected graph, as
described earlier, we see that $\text{nullity}(\widehat\nabla) = E -
(N-1) = \Delta$. Thus the $\Delta$ linearly independent vectors
$\bm{\mathcal{V}}^{(\gamma)}(t)$ span the whole
null space.  If there existed a vector $\delta
{\bm{\mathcal{J}}}(t)$ that satisfied
Eq.~\eqref{f9} but could not be expressed as a linear combination of
basis vectors, then the corresponding vector
$\bm{\mathcal{V}}(t) = \delta
{\bm{\mathcal{J}}}(t) - \delta
{\bm{\mathcal{J}}}^{(1)}(t)$ would be a null
vector that is linearly independent of all the
$\bm{\mathcal{V}}^{(\gamma)}(t)$.  But since the
latter span the whole null space, this is impossible.  Hence every CD
solution $\delta
{\bm{\mathcal{J}}}^{(\gamma)}(t)$ satisfying
Eq.~\eqref{f9} must be expandable in the form of Eq.~\eqref{gs4}.

\subsection{General solution in the cycle basis}\label{sec:cycbasis}

The discussion in the previous section also allows us to rewrite the
expansion in Eq.~\eqref{gs4} in an alternative form that is convenient in practical applications.  Using the fact that
$\sum_{\gamma=1}^{\Delta +1} w_\gamma(t) =1$,
Eq.~\eqref{gs4} can be equivalently expressed as
\begin{equation}\label{gs7}
\begin{split}
    \delta {\bm{\mathcal{J}}}(t) &= \delta {\bm{\mathcal{J}}}^{(1)}(t)+\sum_{\gamma=2}^{\Delta +1} w_\gamma(t) \left(\delta {\bm{\mathcal{J}}}^{(\gamma)}(t)-\delta{\bm{\mathcal{J}}}^{(1)}(t)\right)\\
    &= \delta {\bm{\mathcal{J}}}^{(1)}(t) + \sum_{\gamma=2}^{\Delta +1} w_\gamma(t) {\bm{\mathcal{V}}}^{(\gamma)}(t).
\end{split}
\end{equation}
Since the vectors
${\bm{\mathcal{V}}}^{(\gamma)}(t)$ form a basis
for the null space of $\widehat{\nabla}$, the second term in the last
line of Eq.~\eqref{gs7}, with its arbitrary coefficient functions
$w_\gamma(t)$, is general enough to describe any
vector function in the null space.  With no loss of generality, we can rewrite this second term
in another basis for the null space instead.  A convenient choice is the fundamental cycle basis
corresponding to some reference spanning tree (we need not choose the same reference as used to find
$\delta {\bm{\mathcal{J}}}^{(1)}(t)$).  The
$\Delta$ fundamental cycles were identified in the procedure to
construct derived trees. If we assign an arbitrary orientation to the
cycles (clockwise or counterclockwise), then the $E$-dimensional cycle
vector $\bm{c}^{(\gamma)}$, associated with the derived tree $\gamma +1$, is defined as follows: a $\pm 1$ at every row whose
corresponding edge in the original graph belongs to the fundamental
cycle, with a $+1$ if the edge direction is parallel to the cycle
orientation, $-1$ if anti-parallel.  All edges not belonging to the
fundamental cycle are zero.  For the reference tree in
Fig.~\ref{graphs}C the fundamental cycles are highlighted in green on
the right of the panel.  Here the two cycle vectors are:
\begin{equation}\label{gs7b}
\bm{c}^{(1)} = \begin{pmatrix}
1\\
1\\
1\\
0\\
0
\end{pmatrix}, \qquad 
\rev{\bm{c}^{(2)} = \begin{pmatrix}
1\\
1\\
0\\
1\\
1
\end{pmatrix}}.
\end{equation}
In general, the $\Delta$ fundamental cycle vectors form a basis for the null space of $\widehat{\nabla}$~\cite{deo2017graph}.

In terms of the cycle vectors, Eq.~\eqref{gs7} can be written as
\begin{equation}\label{gs8}
\begin{split}
    \delta {\bm{\mathcal{J}}}(t) =\delta {\bm{\mathcal{J}}}^{(1)}(t) + \sum_{\gamma=1}^{\Delta} v_\gamma(t) \bm{c}^{(\gamma)},
\end{split}
\end{equation}
where $v_\gamma(t)$ for $\gamma =
1,\ldots,\Delta$ are another set of arbitrary coefficient functions.
The convenience of Eq.~\eqref{gs8} over Eq.~\eqref{gs4} is that we
only need to find one spanning tree solution $\delta
{\bm{\mathcal{J}}}^{(1)}(t)$. Both have
the same number of degrees of freedom: in the first case $\Delta$ coefficient
functions $w_\gamma(t)$ for $\gamma
=2,\ldots,\Delta+1$ (since $w_1(t)$ depends on
the rest through the normalization constraint); in the second case
$\Delta$ coefficient functions
$v_\gamma(t)$.

Finally we note that because of Eq.~\eqref{f6}, the oriented
stationary current vector $\bm{\mathcal{J}}(t)$ corresponding
to the original protocol is in the null space of $\widehat\nabla$.
Hence it can also be expanded in terms of the cycle vectors as
\begin{equation}\label{gs9}
\bm{\mathcal{J}}(t) = \sum_{\gamma=1}^{\Delta} u_\gamma(t) \bm{c}^{(\gamma)},
\end{equation}
with some coefficient functions $u_\gamma(t)$.  Since the CD
currents $\bm{\mathcal{\widetilde J}}(t) =
\bm{\mathcal{J}}(t) + \delta
   {\bm{\mathcal{J}}}(t)$, we can combine
   Eqs.~\eqref{gs8} and \eqref{gs9} to get the most general expression
   for any set of currents that satisfies Eq.~\eqref{f8}:
   \begin{equation}\label{gs10}
     {\bm{\mathcal{\widetilde J}}}(t) =\delta {\bm{\mathcal{J}}}^{(1)}(t) + \sum_{\gamma=1}^{\Delta} \Phi_\gamma(t) \bm{c}^{(\gamma)}.     
   \end{equation}
Here $\Phi_\gamma(t) \equiv  u_\gamma(\lambda_t) +  v_\gamma(t)$.  Because the $v_\gamma(t)$ are arbitrary, the functions $\Phi_\gamma(t)$ are also arbitrary, and we still have the same $\Delta$ degrees of freedom to span the solution space.  \rev{Combining Eq.~\eqref{gs10} with Eq.~\eqref{gs3}, we can write the general solution for the CD currents as
\begin{equation}
    \label{sol1}
     {\bm{\mathcal{\widetilde J}}}(t) =[\widehat\nabla^{(1)}]^{-1}_S \partial_t \widehat{\bm{\rho}}(\lambda_t) + C \bm{\Phi}(t),
\end{equation}
where $C$ is the $E \times \Delta$ matrix whose columns are the fundamental cycle vectors $\bm{c}^{(\gamma)}$, $\gamma = 1,\ldots,\Delta$, and $\bm\Phi(t)$ is a $\Delta$-dimensional vector whose components are the arbitrary functions $\Phi_\gamma(t)$.}

\rev{
\subsection{Solving for the CD transition rates}\label{solve}

  Once we know ${\bm{\mathcal{\widetilde J}}}(t)$ from Eq.~\eqref{sol1}, the final step to find the CD protocol is solving for the CD transition rates via Eq.~\eqref{m4ex}.  Using the fact the $\text{diag}(\bm{x}) \bm{y} = \text{diag} (\bm{y}) \bm{x}$ for any vectors $\bm{x}$ and $\bm{y}$, Eq.~\eqref{m4ex} can be rewritten as
\begin{equation}
    \label{sol2}
     {\bm{\mathcal{\widetilde J}}}(t) = M^+(t) \bm{\widetilde{k}}^+(t)  - M^-(t) \bm{\widetilde{k}}^-(t) ,
\end{equation}
where $M^\pm(t)$ is an $E \times E$ diagonal matrix, $M^\pm(t) = \text{diag}({\nabla^\pm}^T \bm{\rho}(\lambda_t))$.  Since every row of ${\nabla^\pm}^T$ has exactly one nonzero element equal to 1, and we focus on systems where the elements of $\bm{\rho}(\lambda_t)$ are all positive, the matrix $M^\pm(t)$ has positive elements on the diagonal and hence is invertible.  We can thus solve Eq.~\eqref{sol2} for $\bm{\widetilde{k}}^+(t)$,
\begin{equation}
    \label{sol3}
    \bm{\widetilde{k}}^+(t) = [M^+(t)]^{-1} {\bm{\mathcal{\widetilde J}}}(t) + [M^+(t)]^{-1} M^-(t) \bm{\widetilde{k}}^-(t).
\end{equation}
Any vectors $\bm{\widetilde{k}}^+(t)$ and $\bm{\widetilde{k}}^-(t)$ with positive elements that satisfy Eq.~\eqref{sol3} constitute valid forward/backward CD transition rates.  If one can control both $\bm{\widetilde{k}}^+(t)$ and $\bm{\widetilde{k}}^-(t)$, valid solutions to Eq.~\eqref{sol3} always exist.

A special case of Eq.~\eqref{sol3} often arises in biological contexts:  for each edge $\alpha$, we can only externally control one rate, which we take to be the forward rate without loss of generality.  The backward rates do not vary in the CD protocol, $\bm{\widetilde{k}}^{-}(t) = \bm{k}^{-}$.  In this scenario, when we use Eq.~\eqref{sol3} to solve for $\bm{\widetilde{k}}^+(t)$, we have to choose the arbitrary function vector $\bm{\Phi}(t)$ in Eq.~\eqref{sol1} to ensure that $\bm{\widetilde{k}}^+(t)$ has positive elements.  If this is not possible then CD driving along the given target trajectory cannot be achieved by only changing the forward rates.

}
\subsection{Thermodynamic costs of CD driving}\label{costs}

\rev{To quantify the thermodynamic costs of driving via the CD protocol, one can calculate the total entropy production rate $\dot{S}^\text{tot}(t)$ at time $t$~\cite{esposito2010},
\begin{equation}\label{cd1}
 \dot{S}^\text{tot}(t) = k_B {\bm{\mathcal{\widetilde J}}}(t) \cdot {\bm{\widetilde \chi}}(t),
\end{equation}
where the $E$-dimensional edge affinity vector ${\bm{\widetilde \chi}}(t)$ is given by
\begin{equation}\label{cd7}
\bm{\widetilde{\chi}}(t) \equiv \ln \left(M^{+}(t) \bm{\widetilde{k}}^+(t)\right)-\ln\left(M^{-}(t)\bm{\widetilde{k}}^-(t)\right).
\end{equation}
Throughout this section we will use the convention that a function like $\ln \mb{v}$ or $\exp \mb{v}$ for a vector $\mb{v}$ is a vector with components $(\ln \mb{v})_\alpha \equiv \ln v_\alpha$ or $(\exp \mb{v})_\alpha \equiv \exp v_\alpha$.  The structure of Eq.~\eqref{cd1}, where every term in the inner product is of the form $(x-y)(\ln x - \ln y)$ for non-negative quantities $x$ and $y$, gives $\dot{S}^\text{tot}(t) \ge 0$, in accordance with the second law of thermodynamics.

Plugging Eq.~\eqref{sol3} into Eq.~\eqref{cd7} we can express $\bm{\widetilde{\chi}}(t)$ as
\begin{equation}\label{t1}
\bm{\widetilde{\chi}}(t) = \ln \left({\bm{\mathcal{\widetilde J}}}(t) + M^-(t) \bm{\widetilde{k}}^-(t)
\right)-\ln\left(M^{-}(t)\bm{\widetilde{k}}^-(t)\right).
\end{equation}
The form of Eq.~\eqref{t1} has an interesting consequence.  Imagine a hypothetical scenario where one can control both the forward $\bm{\widetilde{k}}^+(t)$ and backward $\bm{\widetilde{k}}^-(t)$ rates, satisfying Eq.~\eqref{sol3}.  For a given CD current solution ${\bm{\mathcal{\widetilde J}}}(t)$, the limit where the backward rates become large, $\widetilde{k}_\alpha^-(t) \to \infty$ for all $\alpha$, would correspond to $\bm{\widetilde{\chi}}(t)$ in Eq.~\eqref{t1} becoming arbitrarily small, $|\widetilde{\chi}_\alpha(t)| \to 0$.  Note that due to Eq.~\eqref{sol3} this limit also means 
the forward rates $\widetilde{k}_\alpha^+(t) \to \infty$.  With fixed ${\bm{\mathcal{\widetilde J}}}(t)$ and vanishing $\bm{\widetilde{\chi}}(t)$, we see from Eq.~\eqref{cd1} that the instantaneous entropy production $\dot{S}^\text{tot}(t)$ can approach zero if both backward and forward rates can be made arbitrarily large.

The fact that we can drive the system over a trajectory in finite time with negligible thermodynamic cost is only possible because increasing the rates corresponds to making the local ``diffusivity'' in the system large (if we imagine dynamics on the network as a discrete diffusion process).  In other
words we are reducing the effective friction to zero in order to
eliminate dissipation. In practice this extreme limit is not realistic, particularly for biological systems. There are likely to
be physical constraints that prevent us from simultaneously tuning
each pair of rates in the network over an arbitrary range, so the CD
implementation with $\dot{S}^\text{tot}(t)  \to 0$ is not realizable.

If there is some constraint on at least one of the rates in each pair, the minimum $\dot{S}^\text{tot}(t)$ among all CD protocols for a given target trajectory will have a finite value.  For example, let us take the case mentioned above where the backward rates are fixed, $\bm{\widetilde{k}}^{-}(t) = \bm{k}^{-}$.  Let us also assume that $\Delta >0$, so that varying the functions $\bm{\Phi}(t)$ in Eq.~\eqref{sol1} gives all possible protocols for a particular choice of $\widehat{\bm{\rho}}(\lambda_t)$.  Among these protocols, the condition for minimizing $\dot{S}^\text{tot}(t)$ is given by a gradient with respect to $\bm{\Phi}(t)$ of Eq.~\eqref{cd1},
\begin{equation}
    \label{t2}
    \bm{0} = \bm{\nabla}_{\bm{\Phi}} \dot{S}^\text{tot}(t) = k_B C^T \left(\bm{1} + \bm{\widetilde{\chi}}(t) - e^{-\bm{\widetilde{\chi}}(t)}\right),
\end{equation}
where $\bm{0}$ and $\bm{1}$ denote vectors of zeros and ones respectively (of size $\Delta$ and $E$ in this context).  To derive this, we have used the fact that $\partial \mathcal{\widetilde J}_\alpha(t)/\partial {\Phi_\gamma}  = C_{\alpha\gamma}$ from Eq.~\eqref{sol1}, and that ${\bm{\mathcal{\widetilde J}}}(t) = \Gamma^{-1}(t) \left(e^{\bm{\widetilde{\chi}}(t)} -\bm{1}\right)$ from Eq.~\eqref{t1}, where $\Gamma(t) = [\text{diag}(M^-(t) \bm{k}^- )]^{-1}$.  Note that after finding a set of $\bm{\Phi}(t)$ that satisfies Eq.~\eqref{t2}, one must also check that the corresponding forward CD rates $\bm{\widetilde{k}}^+(t)$ given by Eq.~\eqref{sol3} all have positive elements.

There is one case where Eq.~\eqref{t2} can be solved analytically.  If the driving is slow enough that the magnitudes of the components of ${\bm{\mathcal{\widetilde J}}}(t)$ are small (and hence also those of $\bm{\widetilde{\chi}}(t)$ via Eq.~\eqref{t1}), then $\bm{\widetilde{\chi}}(t) \approx \Gamma(t) {\bm{\mathcal{\widetilde J}}}(t)$.  In this limit Eq.~\eqref{t2} becomes
\begin{equation}
    \label{t3}
    \bm{0} \approx C^T \bm{\widetilde{\chi}}(t) \approx C^T \Gamma(t) {\bm{\mathcal{\widetilde J}}}(t).
\end{equation}
From Eq.~\eqref{t3} and \eqref{sol1} we can now solve for $\bm{\Phi}(t)$,
\begin{equation}
    \label{t4}
    \bm{\Phi}(t) = B(t) \bm{v}(t),
\end{equation}
where $\bm{v}(t) =  [\widehat\nabla^{(1)}]^{-1}_S \partial_t \widehat{\bm{\rho}}(t)$ and the $\Delta \times E$ dimensional matrix $B(t) = -(C^T \Gamma(t) C)^{-1} C^T \Gamma(t)$.  The corresponding minimum instantaneous entropy production for this slow driving regime is:
\begin{equation}
    \label{t5}
    \dot{S}^\text{tot}_\text{min}(t) = \bm{v}^T(t) L^T(t) \Gamma(t) L(t) \bm{v}(t),
\end{equation}
where $L(t) = I_E + C B(t)$, and $I_E$ is the $E \times E$ identity matrix.

The protocol that satisfies Eq.~\eqref{t3} has a particular interpretation:  let us define a $\Delta$-dimensional cycle affinity vector $\bm{\widetilde{\chi}}^\text{cyc}(t) = C^T \bm{\widetilde{\chi}}(t)$, where each component is the sum of the edge affinities over a fundamental cycle~\cite{schnakenberg1976network}.  Equation~\eqref{t3} thus states that $\bm{\widetilde{\chi}}^\text{cyc}(t) = \bm{0}$.  Satisfying these $\Delta$ conditions is equivalent to specifying that the CD protocol rates obey local detailed balance, in other words that there exists a right singular vector of $\widetilde{G}(t)$ with singular value zero.  Hence in the slow driving regime, out of all the possible CD protocols for a given target, it is the one with local detailed balance that minimizes $\dot{S}^\text{tot}(t)$.  This is not generally true outside of the slow driving regime, where Eq.~\eqref{t2} does not coincide with $\bm{\widetilde{\chi}}^\text{cyc}(t) = \bm{0}$.  This observation is compatible with the results of Ref.~\cite{remlein2021optimality}, where they showed among all driving protocols between fixed initial and end distributions the one with the lowest total entropy production violates detailed balance.  Equation~\eqref{t2} shows that this is also true for any specific target trajectory.  Note that when $\Delta =0$ for a system with a tree-like graph, there is no degeneracy in ${\bm{\mathcal{\widetilde J}}}(t)$ and the CD protocol automatically satisfies local detailed balance.
}

\rev{\section{Generalizing counterdiabatic driving:  non-stationary targets and local control}\label{gen}

Up to now the theory has been framed in terms of driving all $N$ states in the system along a target $\bm\rho(\lambda_t)$ that is an instantaneous stationary distribution (Eq.~\eqref{2}) with respect to some original transition matrix $\Omega(\lambda_t)$.  However in biological contexts the control problem may be more general:  perhaps one is interested in having only a subset of states follow  a target trajectory, and the target trajectory does not necessarily have to be a stationary one.  In this section we generalize our theoretical framework to accommodate both these aspects.  Doing so allows us to define a set of graphical rules for controllability in discrete-state Markov models, which we will apply to biological examples in Sec.~\ref{biol}.

\subsection{Driving along non-stationary target trajectories}\label{nonstat}

Let us imagine an arbitrary target trajectory $\bm\rho(t)$ where Eq.~\eqref{2} is not satisfied, and hence $\Omega(\lambda_t) \bm\rho(t) \ne 0$ for at least some $t$ during the driving time interval $0$ to $\tau$.  A special case of this is known as the fast-forward problem, where we start and end in a stationary distribution, $\Omega(\lambda_0) \bm\rho(0) = \Omega(\lambda_\tau) \bm\rho(\tau) = 0$, but allow a non-stationary trajectory for $0 < t < \tau$.  We seek a modified transition matrix $\widetilde{\Omega}(t)$ that satisfies the master equation $\partial_t \bm\rho(t) = \widetilde{\Omega}(t) \bm{\rho}(t)$ during driving.

It turns out that our framework generalizes to arbitrary $\bm\rho(t)$ in a straightforward way:  the main difference is that we no longer enforce Eq.~\eqref{f6}, and we seek a solution to Eq.~\eqref{f8} instead of Eq.~\eqref{f9}.  The general form of $\bm{\mathcal{\widetilde J}}(t)$ that satisfies $\partial_t \bm{\widehat\rho}(t) = \widehat\nabla \bm{\mathcal{\widetilde J}}(t)$ is given by:
\begin{equation}\label{g1}
\bm{\mathcal{\widetilde J}}(t) = [\widehat\nabla^{(1)}]^{-1}_S \partial_t\bm{\widehat{\rho}}(t)+ C \bm{\Phi}^\prime(t).
\end{equation}
Here $[\widehat\nabla^{(1)}]^{-1}_S$ is the stretched inverse corresponding to one of the spanning trees of the graph, and $\bm\Phi^\prime(t)$ is a $\Delta$-dimensional vector whose components are arbitrary functions $\Phi^\prime_\gamma(t)$.  From the fact that $\widehat\nabla [\widehat\nabla^{(1)}]^{-1}_S = I_{N-1}$ and $\widehat\nabla C = \bm{0}$ (since the columns of $C$ are cycle vectors) we see that Eq.~\eqref{g1} does indeed solve Eq.~\eqref{f8}.  Once $\bm{\mathcal{\widetilde J}}(t)$ is known, we find associated transition rates $\bm{\widetilde{k}}^\pm(t)$ that satisfy Eq.~\eqref{sol3}.  In fact, because the structure of Eq.~\eqref{g1} is identical to our earlier CD current expression in Eq.~\eqref{sol1}, the final form of the control protocol solution is the same whether or not the target is stationary.

In order to elucidate the perturbation $\delta \bm{\mathcal{J}}(t)$ to our original currents $\bm{\mathcal{J}}(t)$ necessary achieve the driving, we can write it in the form
\begin{equation}\label{g2}
\begin{split}
\delta \bm{\mathcal{J}}(t) &= \bm{\mathcal{\widetilde J}}(t) - \bm{\mathcal{J}}(t)\\
&= [\widehat\nabla^{(1)}]^{-1}_S\left( \partial_t\bm{\widehat{\rho}}(t)-\widehat\nabla \bm{\mathcal J}(t)\right)+ C \bm{\Phi}(t),
\end{split}
\end{equation}
where the new vector $\bm\Phi(t)$ is defined via
\begin{equation}
    \label{g3}
    C\bm\Phi(t) = C\bm\Phi^\prime(t) + [\widehat\nabla^{(1)}]^{-1}_S \widehat\nabla \bm{\mathcal J}(t) - \bm{\mathcal J}(t).
\end{equation}
The right-hand side of Eq.~\eqref{g3} vanishes when acting on it with $\widehat\nabla$ from the left, and hence it is in the null space of $\widehat\nabla$.  Thus it can be expressed as a linear combination of the fundamental cycle vectors, and hence there must exist a $\bm\Phi(t)$ that satisfies Eq.~\eqref{g3}.  Note that $[\widehat\nabla^{(1)}]^{-1}_S \widehat\nabla$ is an $E \times E$ matrix that does not equal the identity in general, since $[\widehat\nabla^{(1)}]^{-1}_S$ is a right, not left, pseudoinverse of $\widehat\nabla$.  The one case where $[\widehat\nabla^{(1)}]^{-1}_S \widehat\nabla = I_{E}$ is when the original graph is a tree and hence $E = N-1$.  

For the CD driving scenario, where the target trajectory is stationary and $\widehat\nabla \bm{\mathcal J}(t) =0$, Eq.~\eqref{g2} reduces to our earlier CD solution in Eq.~\eqref{gs8}.  More generally, the structure of Eq.~\eqref{g2} allows us to see that only a subset of the currents in the original system need to be modified in order to achieve an arbitrary target.  Since $\bm\Phi(t)$ is arbitrary, we can set $\bm\Phi_\gamma(t) = 0$ for all $\gamma$, and hence from Eq.~\eqref{g2} we get that $\delta \mathcal{J}_{\alpha}(t) \ne 0$ only for those $N-1$ edges $\alpha$ present in the spanning tree (because the rows of $[\widehat\nabla^{(1)}]^{-1}_S$ associated with edges not in the tree are all zero).  If $k^\pm_\alpha(t)$ are the original forward/backward transition rates associated with edge $\alpha$, we have to be able to modify one or both of them to new rates $\widetilde{k}^\pm_{\alpha}(t)$ in order to satisfy the $\delta \mathcal{J}_{\alpha}(t)$ condition, Eq.~\eqref{g2}.  We call an edge $\alpha$ where it is possible to modify the transition rates via external parameters a {\it controllable} edge.  No solution exists with less than $N-1$ controllable edges, and if we set $\bm\Phi_\gamma(t) \ne 0$ we generally get solutions that require more than $N-1$ controllable edges, since the cycle vectors involve currents on edges not in the tree.  Since we can use any spanning tree in Eq.~\eqref{g2}, we can formulate a general rule for {\it global control}, the ability to drive every state in the network along an arbitrary target:\\

{\bf Global control condition:}  in order to drive an $N$-state network along an arbitrary target trajectory $\bm\rho(\lambda_t)$, the set of controllable edges must span the entire network graph.  One consequence is that global control is impossible with less than $N-1$ controllable edges.\\

The minimal condition for global control ($N-1$ controllable edges forming a spanning tree) is the same whether or not the trajectory is an instantaneous stationary one.  Depending on the physical details of a specific system, there may be additional conditions necessary to achieve global control, but the above one must always be fulfilled.  For example, if the backward rates cannot be modified, the forward rates $\bm{\widetilde{k}}^+(t)$ given by Eq.~\eqref{sol3} must all be non-negative. A similar story applies to the case where different controllable edges cannot be independently varied (i.e. because they depend on a single external parameter).  In this situation it may not be possible to simultaneously satisfy the $\delta \mathcal{J}_{\alpha}(t)$ conditions at all controllable edges.

\subsection{Local control}\label{local}

The local control problem means we are only interested in having $N_T < N-1$ of the states in the system follow a target trajectory.  If we label the states such that the first $N_T$ are target states, we need to find $\widetilde{\Omega}(t)$ such that the master equation $\partial_t \bm{p}(t) = \widetilde{\Omega}(t) \bm{p}(t)$ is satisfied with a solution whose probability vector has the form
\begin{equation}
    \label{g4}
    \begin{split}
    \bm{p}(t) &= \left(\rho_1(t), \ldots,\rho_{N_T}(t), \pi_{N_T+1}(t),\ldots,\pi_{N}(t)\right)\\
    &\equiv (\bm{\rho}(t), \bm{\pi}(t)),
    \end{split}
\end{equation}
where we denote the $N_T$-dimensional target trajectory vector $\bm{\rho}(t)$ and the $(N-N_T)$-dimensional vector of remaining non-target states as $\bm{\pi}(t)$.  As discussed below, the case where $N_T = N-1$ corresponds to the global control scenario, since the probability of the $N$th state is constrained by the normalization condition $\sum_{i=1}^N p_i(t)=1$.  The target trajectory $\bm{\rho}(t)$ is specified beforehand, and we are looking for all possible solutions compatible with a given $\bm{\rho}(t)$.

\begin{figure}
    \centering
    \includegraphics[width=\columnwidth]{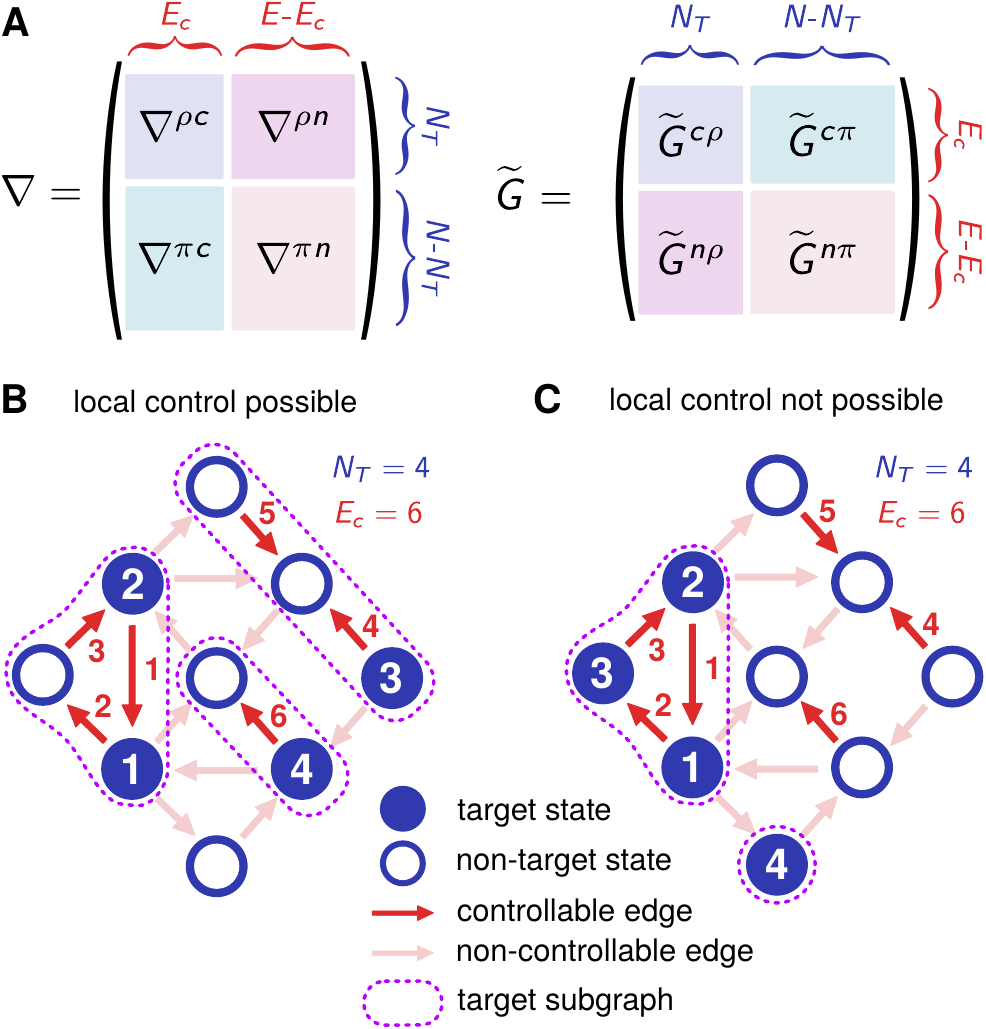}
    \caption{\rev{A) Partition of the graph matrices $\nabla$ and $\widetilde{G}(t)$ into submatrices to facilitate solving the local control problem.  B) A network with $N_T = 4$ target states and $E_C = 6$ controllable edges, depicted as indicated by the legend.  The three target subgraphs, each containing at least one target state and all other states connected to it via controllable edges, are outlined in dashed curves.  Because each target subgraph includes at least one non-target state, the local control condition is satisfied.  C) Same network as B, except that the set of $N_T = 4$ target states is different.  Here neither of the two target subgraphs contain any non-target states, and hence local control is impossible.}}
    \label{control}
\end{figure}

Imagine there are $E_c \le E$ edges in the system that are controllable, and we label the edges such that $\alpha = 1,\ldots,E_c$ correspond to the controllable ones.  We can then partition the current vector $\bm{\widetilde{\mathcal J}}(t)$ into controllable (supserscript c) and not-controllable (superscript n) components as follows:
\begin{equation}
    \label{g5}
    \begin{split}
    \bm{\mathcal{\widetilde J}}(t) &= \left(\mathcal{\widetilde J}^c_1(t), \ldots,\widetilde{\mathcal J}^c_{E_c}(t), \widetilde{\mathcal J}^n_{E_c+1}(t), \ldots,\widetilde{\mathcal J}^n_{E}(t)\right)\\
    &\equiv (\bm{\mathcal{\widetilde J}}{}^c(t), \bm{\mathcal{\widetilde J}}{}^n(t)).
    \end{split}
\end{equation}
Analogously, we can partition the forward/backward rates on the edges:
\begin{equation}
    \label{g5b}
    \begin{split}
    \bm{\widetilde{k}}^\pm(t) &= \left({\widetilde{k}}_1^{\pm c}(t),\ldots,{k}_{E_c}^{\pm c}(t),{k}_{E_c+1}^{\pm n},\ldots,{\widetilde{k}}_{E}^{\pm n}\right)\\
    &\equiv (\bm{\widetilde k}^{\pm c}(t), \bm{k}^{\pm n}).
    \end{split}
\end{equation}
Note that by definition the rates $\bm{k}^{\pm n}$ on the non-controllable edges cannot be externally modified. We additionally assume here that they are time-independent, the typical case for biological systems.  This allows us to find the analytical solution for $\bm{\pi}(t)$ shown below.  If $\bm{k}^{\pm n}(t)$ are fixed, time-dependent functions, Appendix~\ref{deriv} shows how the general solution requires numerically solving a differential equation for $\bm{\pi}(t)$.  Finally, we can partition the graph matrices $\nabla$ and $\widetilde{G}(t)$ each into four submatrices as shown in Fig.~\ref{control}A.

  At the outset of the local control problem, the following quantities are known:  the target trajectory $\bm{\rho}(t)$, the non-controllable edge rates $\bm{k}^{\pm n}$, the four submatrices of $\nabla$, and the bottom two $\widetilde{G}(t)$ submatrices, $\widetilde{G}^{n\rho}$ and $\widetilde{G}^{n\pi}$ (which depend only on the non-controllable rates, and hence are time-independent).  The goal is to figure out controllable edge rates $\bm{\widetilde{k}}^{\pm c}(t)$ that force to system to follow the trajectory on the target states.  A detailed derivation of the solution is presented in Appendix~\ref{deriv}.  Here we summarize the resulting procedure, which consists of three steps.

\vspace{0.5em}
\noindent {\bf i)} Solve for $\bm{\pi}(t)$:
 \begin{equation}\label{gex1}
 \bm{\pi}(t) = e^{t B \widetilde{G}^{n\pi}} \bm{\pi}(0) + \int_0^t dt^\prime\, e^{(t-t^\prime) B \widetilde{G}^{n\pi} } \bm{a}(t^\prime),
 \end{equation}
 where
 \begin{equation}\label{gex2}
     \begin{split}
         B &= \nabla^{\pi n} - \nabla^{\pi c} [\nabla^{\rho c}]^{-1}_S \nabla^{\rho n},\\
         \bm{a}(t) &= \nabla^{\pi c}[\nabla^{\rho c}]^{-1}_S \partial_t \bm{\rho}(t) + B \widetilde{G}^{n\rho} \bm{\rho}(t) + \nabla^{\pi c} C^c \bm{\Phi}^c(t).
     \end{split}
 \end{equation}
Here $\bm{\pi}(0)$ is a set of arbitrary initial probabilities for the non-target states, with the constraint that the components of $\bm{\pi}(0)$ and $\bm{\rho}(0)$ must add up to 1.  $C^c$ is a matrix with dimensions $E_c \times \Delta_c$ whose columns form a basis for the null space of $\nabla^{\rho c}$.  As we will argue below, $\Delta_c = E_c - N_T$ when a local control solution exists. $\bm{\Phi}^c(t)$ is a $\Delta_c$-dimensional vector of arbitrary functions.  The freedom to choose $\bm{\pi}(0)$ and $\bm{\Phi}^c(t)$ means that in general the solution for $\bm{\pi}(t)$ is non-unique.

\vspace{0.5em}
\noindent {\bf ii)} Solve for the currents at the controllable edges:
\begin{equation}
    \label{gex3}
    \begin{split}
\bm{\mathcal{\widetilde J}}{}^c(t) =& [\nabla^{\rho c}]^{-1}_S \left(\partial_t\bm{\rho}(t)  - \nabla^{\rho n} \widetilde{G}^{n\rho} \bm{\rho}(t) - \nabla^{\rho n}\widetilde{G}^{n\pi}\bm{\pi}(t)\right)\\ &+ C^c \bm{\Phi}^c(t).
\end{split}
\end{equation}

\vspace{0.5em}
\noindent {\bf iii)} Solve for the controllable edge rates:
\begin{equation}
    \label{gex4}
    \bm{\widetilde{k}}^{+ c}(t) = [M^{+c}(t)]^{-1} {\bm{\mathcal{\widetilde J}}}{}^c(t) + [M^{+c}(t)]^{-1} M^{-c}(t) \bm{\widetilde{k}}^{-c}(t),
\end{equation}
where $M^{\pm c}(t) = \text{diag}({\nabla^{\pm \rho c}}^T \bm{\rho}(t) + {\nabla^{\pm \pi c}}^T \bm{\pi}(t))$.  Here ${\nabla^{\pm \rho c}}$ and ${\nabla^{\pm \pi c}}$ refer to submatrices of $\nabla^\pm$ of the same form as those for $\nabla$ in Fig.~\ref{control}A.

As a consistency check, we note that the above approach recovers our earlier global control results in the limit when $N_T = N-1$ and $E_c = E$.  In this case $\nabla^{\rho c} = \widehat{\nabla}$, $\nabla^{\rho n} = \nabla^{\pi n} = 0$, $\Delta_c  = \Delta$, $C^c =C$, and $[\nabla^{\rho c}]^{-1}_S = [\widehat{\nabla}^{(1)}]^{-1}_S$.  Because $\nabla^{\pi c}$ here is the last row of $\nabla$, and each column of $\nabla$ sums to zero, $\nabla^{\pi c}$ is just minus the sum of the rows of $\widehat\nabla$.  This implies that $\nabla^{\pi c} [\nabla^{\rho c}]^{-1}_S = (-1,\ldots,-1)$ and $\nabla^{\pi c} C^c =0$.  Hence Eq.~\eqref{gex2} simplifies to $B = 0$ and $a(t) = -\sum_{i=1}^{N-1} \partial_t \rho_i(t)$, so that Eq.~\eqref{gex1} becomes $\pi_N (t) = 1 - \sum_{i=1}^{N-1} \rho_i(t)$.  In a similar way, Eq.~\eqref{gex3} becomes Eq.~\eqref{g1} and Eq.~\eqref{gex4} becomes Eq.~\eqref{sol3}.

A local control solution is not always possible, since there are two criteria that need to be satisfied.  The first is that there must exist an $E_c \times N_T$ stretched inverse matrix $[\nabla^{\rho c}]^{-1}_S$ such that $\nabla^{\rho c} [\nabla^{\rho c}]^{-1}_S  = I_{N_T}$.  The second is that the components $\bm{\pi}(t)$ from Eq.~\eqref{gex1} need to be valid probabilities, $\pi_i(t) \ge 0$ for all $t$ during the driving protocol.  Note that normalization, where the components of $\bm{\pi}(t)$ and $\bm{\rho}(t)$ sum to 1 at all $t$, is guaranteed by the structure of the solution, but $\pi_i(t) \ge 0$ is not automatically enforced.  Given the ability to choose $\bm{\pi}(0)$ and $\bm{\Phi}^c(t)$, it is often feasible to satisfy this second criterion.  For the first criterion to work, $\nabla^{\rho c}$ must have rank $N_T$, and there is a simple graphical method to check for this.  Let us define a {\it target subgraph} as follows:  starting from any state in the target subset, this is the connected subgraph of all states reachable via only controllable edges.  Examples of target subgraphs are highlighted with dashed curves in Fig.~\ref{control}B,C.  These two panels depict the same network and same set of $E_c =6$ controllable edges, but with two different sets of target states with $N_T =4$.  There may be multiple target subgraphs in a network, involving disjoint subsets of the controllable edges.  As shown in the figure, a target subgraph must include at least one target state, but can also include non-target states.  We can now formulate the general rule for local control, encapsulating both criteria:\\

{\bf Local control condition:} in order to drive $N_T < N-1$ states from a network of size $N$ through an arbitrary trajectory $\bm{\rho}(t)$, every target subgraph must include at least one non-target state.  One consequence is that local control is impossible with less than $N_T$ controllable edges.  An additional criterion is that there must be a solution for $\bm{\pi}(t)$ with non-negative components during the time interval of driving.\\

The reason that the subgraph condition is sufficient for $[\nabla^{\rho c}]^{-1}_S$ to exist is that we can choose one of the non-target states in each target subgraph as the analogue of a reference state for that subgraph.  This makes the row of $\nabla^{\rho c}$ for each target state equal to the row of the reduced incidence matrix for the target subgraph to which the state belongs.  We know that rows of a reduced incidence matrix are linearly independent from one another for the same subgraph, and for different subgraphs they are linearly independent because they involve different subsets of edges.  Thus overall if the local control condition is satisfied, all rows of $\nabla^{\rho c}$ are linearly independent.  The condition also implies a lower bound on the number of controllable edges, $E_c \ge N_T$.  To see this, let us imagine there are $K$ target subgraphs in the network, $\kappa = 1,\ldots,K$ each containing $n_\kappa$ target states and at least one non-target state.  For each subgraph to be connected, it must involve at least $n_\kappa$ controllable edges.  We thus get that $E_c \ge \sum_{\kappa=1}^K n_\kappa = N_T$.  The linear independence of the $N_T$ rows of $\nabla^{\rho c}$, plus the fact that the number of columns $E_c$ is at least $N_T$, guarantees that $\nabla^{\rho c}$ has rank $N_T$.  Since the rank and nullity of $\nabla^{\rho c}$ must sum to $E_c$, this also implies that its nullity $\Delta_c = E_c - N_T$.

When the local control condition is satisfied, we can write down the stretched inverse $[\nabla^{\rho c}]^{-1}_S$ using a graphical procedure analogous to the global control case.  First, choose a controllable edge spanning tree for each target subgraph.  To find the $i$th column of $[\nabla^{\rho c}]^{-1}_S$, follow the tree path from the non-target reference state to state $i$ in the corresponding subgraph, and put a $\pm 1$ at each row $\alpha$ where the corresponding edge is parallel / anti-parallel to the path.  All other entries in the column are zero.
}

\section{Driving in biological networks}\label{biol}
% \section{CD driving in biological discrete state models}\label{biol}

To illustrate the general theory in specific biological contexts, we consider two examples of driving in biochemical networks, \rev{corresponding to global and local control respectively}. The first example is a simple genetic regulatory switch involving a repressor protein and corepressor ligand binding to an operator site on DNA, turning off the expression of a set of genes.
Here it turns out there are enough control knobs---concentrations of
repressors, corepressors, and repressor-corepressor complexes---to
implement a whole family of exact \rev{global control} solutions.  Among this family we
can then examine which ones satisfy certain physical
constraints, or minimize thermodynamic costs.  The second
example involves a chaperone protein that binds to a misfolded
substrate, catalyzing the unfolding of this misfolded
protein and giving it another chance to fold into the
correct (``native'') state.  The available
control knobs---chaperone and ATP concentrations--\rev{are insufficient for global control, but do allow the system to locally control the probability of being in the misfolded state.  This local control turns out to be of crucial importance, since rapidly decreasing the misfolded probability is a way to ameliorate the damage due to heat shock.  In fact the local control protocols from our theory qualitatively mimic experimental results from yeast and {\it E. coli}.}

\subsection{Repressor-corepressor model}

\begin{figure}
\includegraphics[width=\columnwidth]{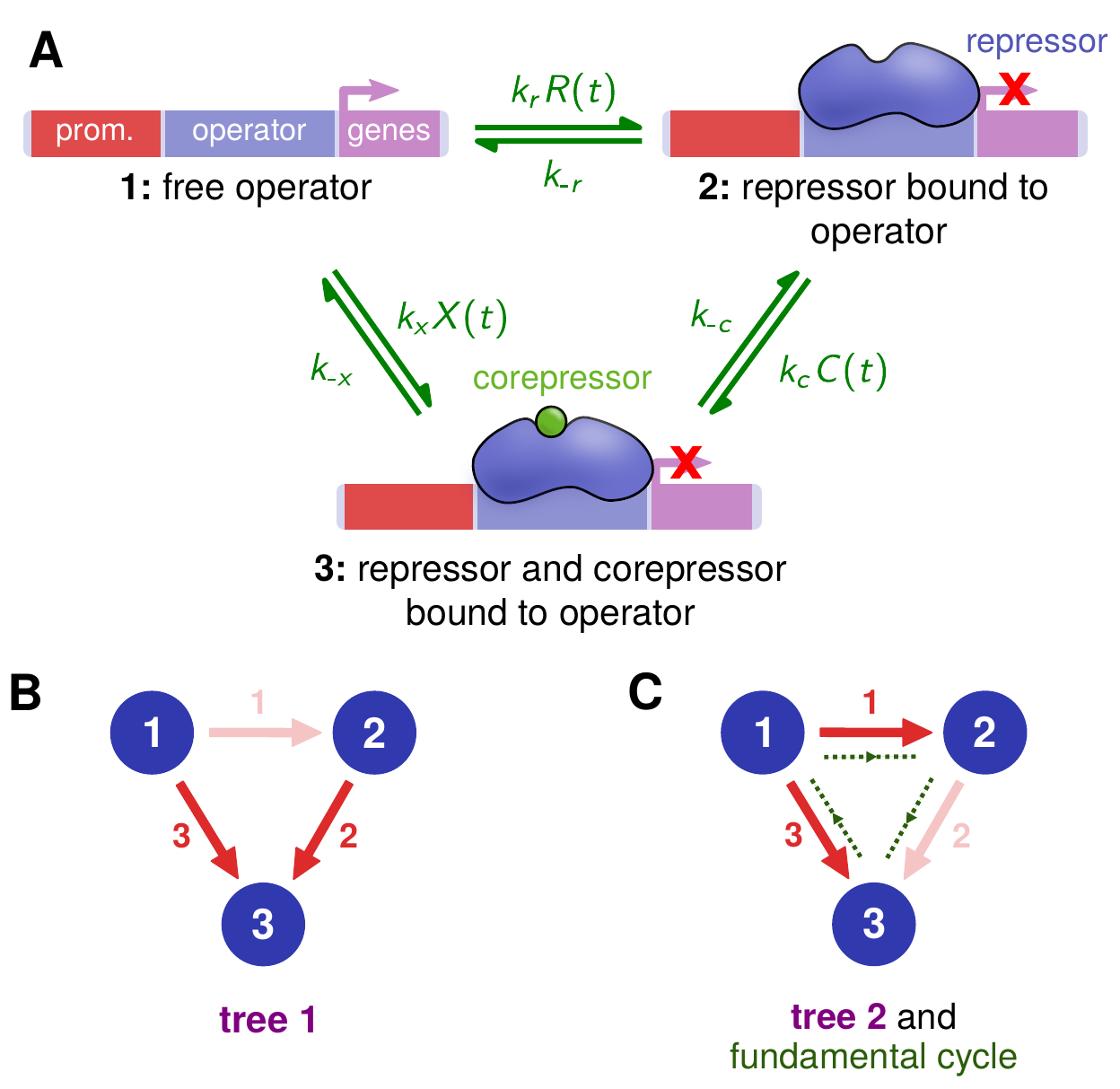}
\caption{A) Biochemical network of a repressor-corepressor model, showing an operator site on DNA in three different states: 1) free; 2) bound to a bare repressor protein; 3) bound to a represssor-corepressor complex.  Transition rates between the states are shown in green.  The binding reaction rates depend on three concentrations of molecules in solution: $R(t)$ for bare repressors, $C(t)$ for corepressors, and $X(t)$ for the complexes.  B) One of the spanning trees for the associated network graph, with the edge deleted to form the tree shown in faint red.  We take this to be the reference spanning tree for the tree basis.  C) The other tree in the basis, with the corresponding fundamental cycle in green.}\label{rep}
\end{figure}

The first system we consider is a common form of gene regulation in
bacteria, illustrated schematically in Fig.~\ref{rep}A: a repressor
protein has the ability to bind to an operator site on DNA.  When
bound, it interferes with the ability of RNA polymerase to attach to
the nearby promoter site, preventing the transcription of the genes
associated with the promoter.  The system acts as a genetic
switch, with the empty operator site the ``on'' state for gene
expression, and the occupied operator site the ``off'' state.  In many
cases, additional regulatory molecules---inducers or
corepressors--- influence the binding affinity of repressor proteins~\cite{schumacher1995mechanism}.  In the present model,
binding of the bare repressor to the operator site is weak (it unbinds
easily), but the binding strength is enhanced in the presence of a particular small molecule---the corepressor.  Hence the
corepressor concentration acts like an input signal, with sufficiently
high levels leading to the promoter site being occupied with high
probability, and the associated genes being turned off.  \rev{Such genetic switches are basic building blocks of natural and synthetic biological circuits.  From the control standpoint, can we drive the switch through a prescribed trajectory, turning it on or off in a finite time, and at what cost?}

\rev{There are several reasons this system provides a convenient testing ground for our theory.  As described below, it can be modeled with three discrete states connected via Markovian transitions, forming a three-state loop (Fig.~\ref{rep}A).  This is the simplest graph structure where there exists a whole family of global control protocols for any given target trajectory.  Though each of these protocols achieves the same target, they are chemically and thermodynamically distinct, allowing us to explore interesting facets of degeneracy in the driving theory.  Since all the forward and reverse rates of the system are known experimentally, taken from {\it in vitro} measurements of the purine repressor (PurR) system of {\it
  E. coli}~\cite{schumacher1995mechanism,xu1998kinetic}, the system also provides a simple platform to directly test theoretically predicted control protocols in the future (for example using a time-resolved version of the {\it in vitro} fluorescence spectroscopy already successfully applied to PurR in Ref.~\cite{xu1998kinetic}).  Finally, the general structure of the network, with edges where either the forward or backward transition depends on the concentration of a regulatory molecule or enzyme, is quite representative of biochemical systems in general.  Hence it serves as a jumping off point for the analysis of more complex biological networks.}

\begin{figure}
  \includegraphics[width=\columnwidth]{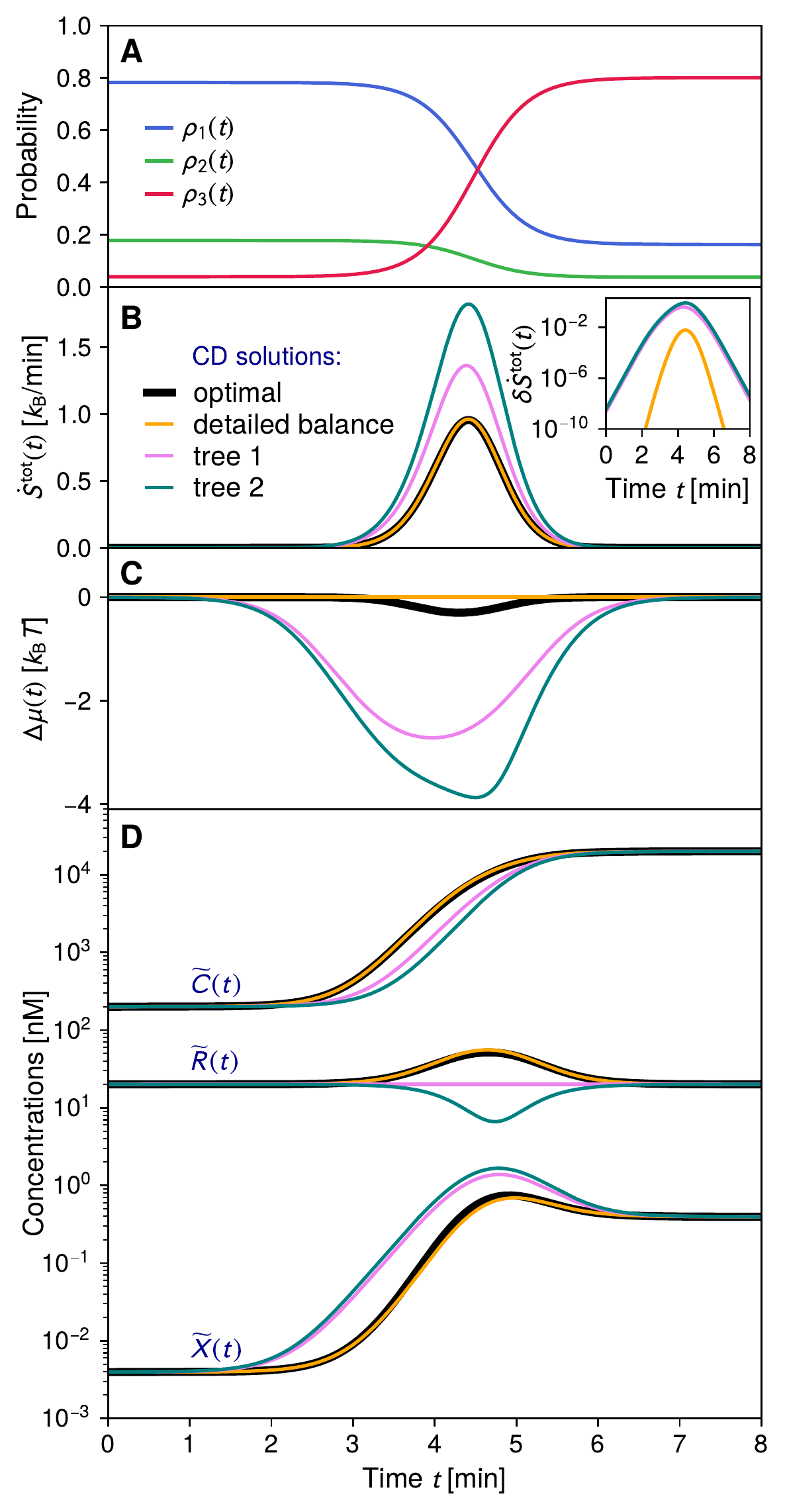}
\caption{A) Components of the target stationary distribution trajectory $\bm{\rho}(\lambda_t)$ (solid curves) for the repressor-corepressor system. B-D) Characteristics of four different \rev{control protocols} that all drive the system along the target trajectory $\bm{\rho}(\lambda_t)$.  The four \rev{protocols} are: spanning tree 1 (violet, corresponding to Fig.~\ref{rep}B); spanning tree 2 (teal, corresponding to Fig.~\ref{rep}C); the solution satisfying local detailed balance, $\Delta\mu(t)= 0$ at all $t$ (yellow); and the optimal solution that minimizes $\dot{S}^\text{tot}(t)$ at all $t$ (thick black).  For each \rev{case} we depict:  B) the total entropy production rate $\dot{S}^\text{tot}(t)$, with the inset showing the difference $\delta \dot{S}^\text{tot}(t) \equiv \dot{S}^\text{tot}(t)-\dot{S}^\text{tot,opt}(t)$ between non-optimal and optimal rates on a log scale [units: $k_B$/min]; C) the instantaneous chemical potential $\Delta\mu(t)$; D) the concentrations $\widetilde{C}(t)$, $\widetilde{R}(t)$, $\widetilde{X}(t)$.}\label{repd}
\end{figure}

As is generally the case with genetic regulation in biology, the
processes underlying repressor dynamics are
stochastic~\cite{berg2000fluctuations}.  \rev{The three discrete states in our Markov model are}: 1) free
operator; 2) bare repressor bound to the operator; 3)
repressor-corepressor complex bound to the operator.  Transitions in
both directions (clockwise and counterclockwise in Fig.~\ref{rep}A)
are possible.  In each pair of transition rates between neighboring
states there is a binding reaction proportional to the
concentration of a chemical species in solution.
% in the environment of the system.  
The relevant concentrations are those of bare repressors
$R(t)$, corepressors $C(t)$, and repressor-corepressor complexes
$X(t)$.  \rev{If we label the binding reactions as the forward rates, then $\bm{k}^+(t) = (k_r R(t), k_c C(t), k_x X(t))$, with associated binding constants $k_\mu$ for each species $\mu$.  The concentration dependence of the forward rates means we have three controllable edges, thus satisfying the global control criterion (at least 2 controllable edges for a 3-state network).  The reverse rates, describing the unbinding reactions, cannot be externally tuned: $\bm{k}^- = (k_{-r}, k_{-c}, k_{-x})$.}

To be concrete, we \rev{choose parameters} based on
the purine repressor (PurR) system of {\it
  E. coli}~\cite{schumacher1995mechanism,xu1998kinetic}.  The PurR
protein turns off genes responsible for the \emph{de novo} production of
purines, a class of molecules including guanine and adenine that are
essential ingredients in DNA/RNA and energy transducing molecules like ATP.  If
the cell has an excess of purines (for example from environmental
sources), this is signaled by an abundance of the corepressors guanine
or hypoxanthine (a purine derivative) that form complexes with
PurR, enabling it to bind strongly with its operator site.  This way,
the cell can switch off the energetically expensive \emph{de novo} production
of purines when it is not needed.  \rev{The parameter values, as well as the calculation of the control protocols, are shown in Appendix~\ref{repdet}.

For a given target trajectory $\bm{\rho}(t)$, the goal is to find forward rates  $\bm{\widetilde{k}}^+(t) = (k_r \widetilde{R}(t), k_c \widetilde{C}(t), k_x \widetilde{X}(t))$ that force the system to be on-target.  These define concentration protocols $\widetilde{R}(t)$, $\widetilde{C}(t)$, and $\widetilde{X}(t)$ for the three species.  Fig.~\ref{repd}A shows our chosen target $\bm{\rho}(t)$, mimicking a biological scenario where the genetic switch is rapidly turned off over the course of a couple of minutes:  the free operator (state 1) probability is decreased, with a corresponding increase in the repressor-bound states 2 and 3.  This particular $\bm{\rho}(t)$ (details in Appendix~\ref{repdet}) consists of instantaneous stationary distributions for the system, so the driving is CD, but as described in Sec.~\ref{nonstat} any other $\bm{\rho}(t)$ could have been chosen.

Since $\Delta = 1$ for the oriented current graph, there will be many possible concentration protocols that drive the system along exactly the same target, via different choices of $\Phi_1(t)$ in Eq.~\eqref{sol1} for the CD currents.  Fig.~\ref{repd}D shows four different examples of concentration protocols.  The violet and teal curves are for the tree 1 and 2 solutions (Fig.~\ref{rep}B,C) where $\mathcal{\widetilde{J}}_1(t) = 0$ and $\mathcal{\widetilde{J}}_2(t) = 0$ respectively.  The other two protocols are described below.  Despite leading to the same system behavior, the protocols have quite distinct physical characteristics, with concentrations varying up to an order of magnitude among the four examples shown.  They also differ in the cycle affinity $\widetilde{\chi}^\text{cyc}(t)$, which is a scalar since $\Delta =1$.  This affinity has a more direct physical interpretation as the chemical potential $\Delta \mu(t) = k_B T \widetilde{\chi}^\text{cyc}(t)$ for the repressor-corepressor binding reaction, and is plotted in Fig.~\ref{repd}C for the four protocols.  One of the protocols (yellow curve) has rates chosen so that $\Delta \mu(t) = 0$, satisfying the local detailed balance condition.  As described in Sec.~\ref{costs}, we know that this protocol should be the one with the smallest entropy production rate in the slow driving limit.  Here we are away from that limit, but the protocol still does well in economizing thermodynamic costs, as seen in the plot of $\dot{S}^\text{tot}(t)$ in Fig.~\ref{repd}.  The protocol that satisfies Eq.~\eqref{t2}, and hence optimizes $\dot{S}^\text{tot}(t)$, is shown as a thick black curve for comparison.  It is close, but not exactly equal to, the detailed balance protocol, exhibiting slightly negative $\Delta\mu(t)$ at intermediate times (Fig.~\ref{repd}C).  The inset of Fig.~\ref{repd}B shows the difference $\delta\dot{S}^\text{tot}(t)\equiv \dot{S}^\text{tot}(t)-\dot{S}^\text{tot,opt}(t)$ between each non-optimal solution and the optimal one.  The detailed balance solution is significantly closer to optimal entropy production than the two tree solutions.

The repressor-corepressor model illustrates the variety of physically realizable control solutions that can exist in certain cases.  This gives nature (or an experimentalist engineering a synthetic system) a rich array of options to achieve a specific probabilistic target.  When we observe, for example, a genetic switch within a biological circuit being rapidly turned off by changing concentrations of external species, there will typically be a variety of alternatives that would have led to same state distribution at each instant of time.  An interesting question for future studies would be to ask whether certain options would be evolutionarily favored over others because of selection pressures due to energetic costs~\cite{ilker2019}.}

\subsection{Chaperone model}\label{chapmodel}

Many newly synthesized proteins, susceptible to misfolding, become
trapped in long-lived metastable states that are prone to aggregation.
Since aggregates present a danger to the survival of the cell, there
exists an elaborate rescue machinery of molecular chaperone proteins
that facilitate unfolding or disaggregating misfolded
proteins~\cite{lorimer1996,thirumalai2001,kerner2005,santra2017}.  In
the case of ${\it E. coli}$, which has the most extensively studied
chaperone network, certain components like the GroEL-GroES system are
obligatory for survival~\cite{fayet1989}.  Environmental stresses
further exacerbate the problem, and an increase of ambient
temperature by just a few degrees can significantly enhance protein
misfolding and consequently aggregation~\cite{richter2010}.  Responding to a heat shock requires creating extra capacity, since
even under normal conditions the majority of chaperones are occupied
by misfolded proteins~\cite{kerner2005} (i.e. occupancy for GroEL can
approach 100\% for fast-growing {\it E. coli}~\cite{santra2017}).  This
is accomplished by rapidly upregulating the number of chaperones to cope with
additional misfolded proteins~\cite{richter2010,roncarati2017}.

Most chaperones require constant power input in the form of ATP
hydrolysis. As a result the stationary probability distribution of
conformational states for a protein interacting with a chaperone will
generally be out of equilibrium
(non-Boltzmannian)~\cite{chakrabarti2017,goloubinoff2018}.  When the
chaperone concentration increases after a heat shock (for example
following a sudden rise to a new temperature~\cite{soini2005}), the
protein is driven away from the previous stationary distribution, and eventually relaxes to a new stationary distribution once the chaperone
concentrations reach steady-state values at the new temperature.
Chaperone upregulation during heat shock therefore serves as a natural
example of nonequilibrium driving in a biological system.

\begin{figure}[t]
\includegraphics[width=\columnwidth]{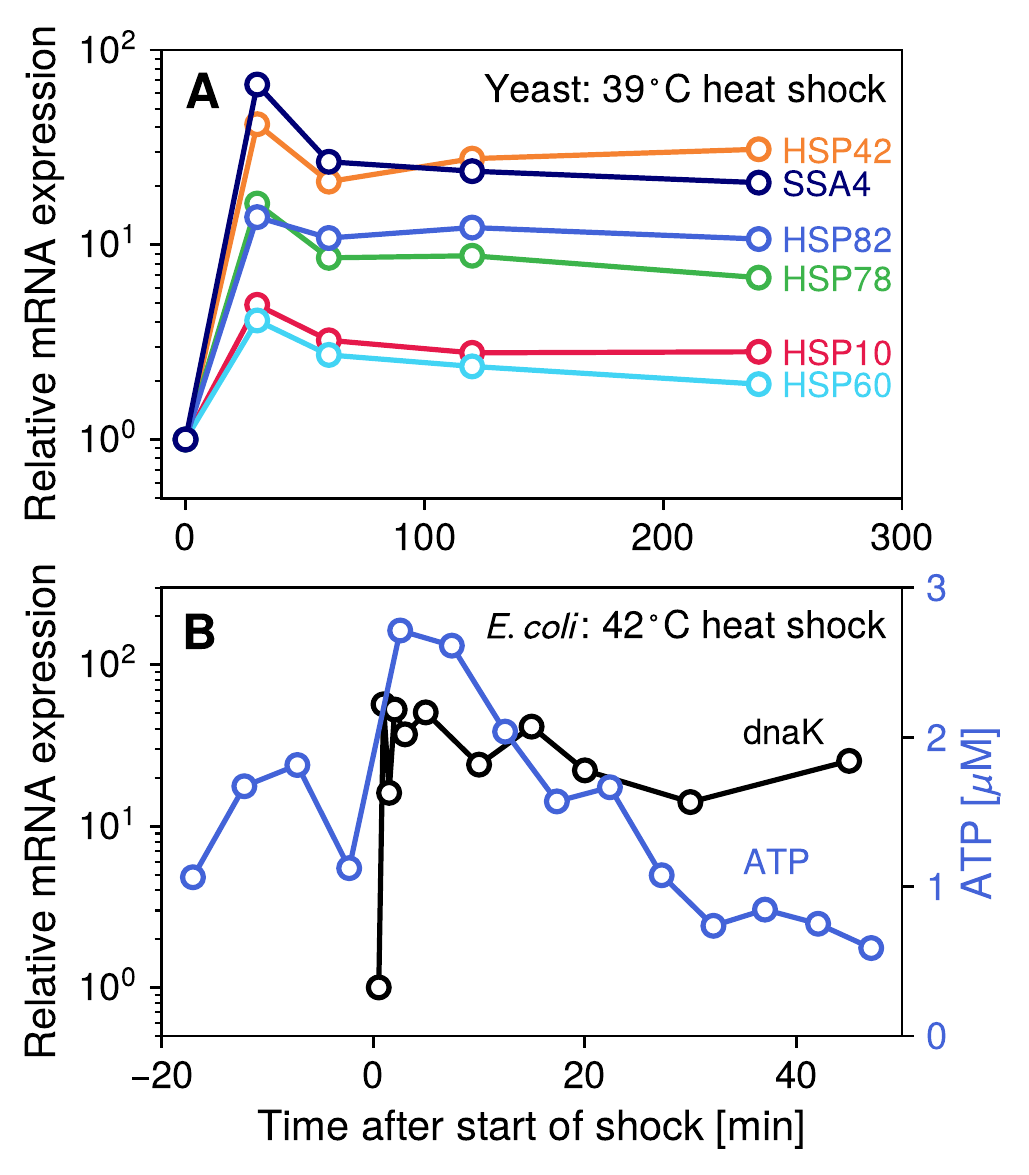}
\caption{\rev{Experimental results for the heat shock response of yeast and {\it E. coli}.  A) Relative mRNA expression of six chaperone genes in {\it S. cerevisiae} yeast versus time after a the start of a $30^\circ \to 39^\circ$C heat shock.  The genes, listed on the right, all have the Gene Ontology  database annotation 0051082~\cite{ashburner2000gene,gene2021gene}, indicating that their products exhibit chaperone activity (binding to unfolded proteins). B) The black curve shows the relative mRNA expression of the {\it E. coli} chaperone gene dnaK versus time after the start of a $30^\circ \to 42^\circ$C heat shock~\cite{soini2005}.  The blue curve shows ATP concentration for the same system.}}\label{exp}
\end{figure}

\begin{figure*}
\includegraphics[width=\textwidth]{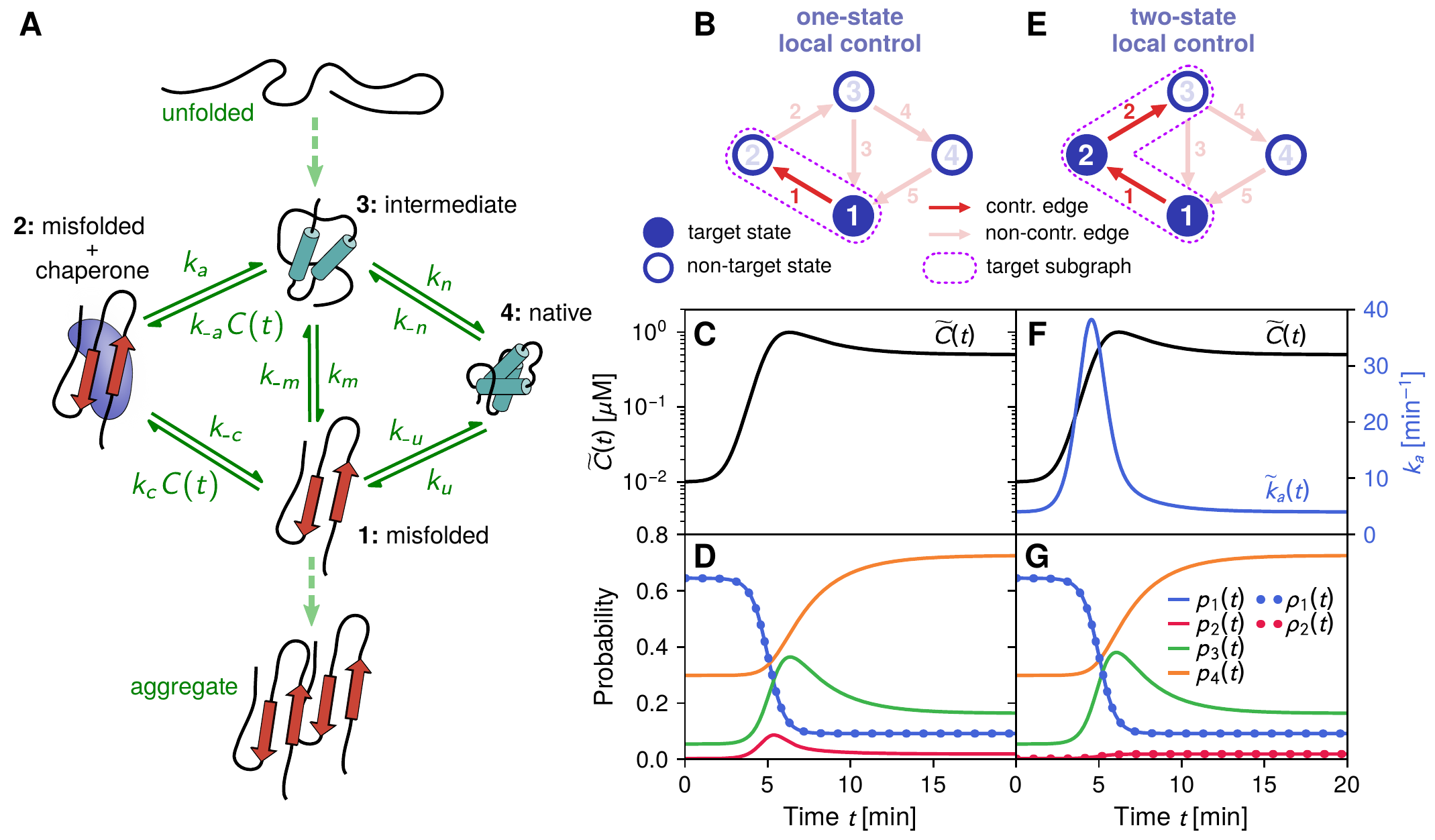}
\caption{A) Conformational states of a protein interacting with a chaperone.  Transition rates in our kinetic network model are indicated by solid green arrows.  Related transitions outside the scope of the model are shown as dashed arrows.  \rev{B) If $k_a$ is fixed and $k_{-a} C(t)$ is negligible, there is effectively only one controllable edge in the network, the one between state 1 and 2.  We can thus choose a target subgraph to enable one-state local control, with state 1 (misfolded) as our target.  C-D) Example of a one-state local control solution, with a chaperone concentration protocol $\widetilde{C}(t)$ in panel C forcing the state 1 probability $p_1(t)$ to follow a target sigmoidal decrease in the misfolded probability $\rho_1(t)$ in panel D.  The state probabilities $p_i(t)$, $i=1,\ldots,4$ are calculated by numerical solution of the master equation, with parameters described in Appendix~\ref{chapdet}.  E-G) Same as panels B-D except with an additional controllable edge due to varying ATP levels, allowing $k_a(t)$ to be time-dependent.  We can now have two-state local control, targeting states 1 and 2, and find a protocol of $\widetilde{C}(t)$ and $\widetilde{k}_a(t)$ to make the system follow a chosen set of targets $\rho_1(t)$ and $\rho_2(t)$.}}
\label{f1}
\end{figure*}

\rev{Fig.~\ref{exp} shows experimental results for the heat shock response of two representative organisms.  In Fig.~\ref{exp}A relative mRNA levels for six different chaperone genes in {\it S. cerevisiae} yeast are plotted as function of time~\cite{truttmann2017unrestrained}.  Since higher mRNA expression generally leads higher concentrations of the proteins coded for by the mRNA, the mRNA levels can be seen as a proxy for chaperone concentration.  At the start of the experiment, the temperature is raised from 30$^\circ$ to 39$^\circ$C, and then held constant.  Chaperone gene expression rises sharply in the first half-hour (in some cases by more than an order of magnitude), then peaks and levels off at a value roughly half that of the peak.  {\it E. coli} shows similar behavior (Fig.~\ref{exp}) for mRNA levels of the dnaK chaperone gene~\cite{soini2005}, in this case following a heat shock from 30$^\circ$ to 42$^\circ$C.  In both organisms the chaperone levels overshoot and then remain elevated for a long duration after the shock, a characteristic feature of the heat shock response~\cite{eisen1998,richter2010}.  Interestingly, {\it E. coli} shows another, less common, behavior:  ATP concentration transiently increases by about a factor of two in the first minutes after the shock, an observation additionally supported by metabolic evidence~\cite{angles2017}.  How do such changes in chaperones and ATP affect the state distribution of a protein targeted by chaperones?  In the analysis below, we will see that these two control knobs enable local control of states involving the misfolded protein.

Our starting point is} a four-state Markov model for chaperone-assisted protein unfolding, inspired by earlier models like those of Refs.~\cite{chakrabarti2017,goloubinoff2018}.  We focus on a network of four states for a particular substrate (``client'') protein, and one type of chaperone, depicted in Fig.~\ref{f1}A:  \rev{1)} a misfolded protein state, prone to aggregation; 2) the misfolded protein bound to chaperone; \rev{3)} an intermediate conformational state of the protein, along the folding pathway between the unfolded and native states; 4) the correctly folded ``native'' state.  These four states can interconvert with transition rates denoted in the figure (further details below).  The model is a small biochemical module within a broader set of processes, some of which are depicted schematically with dashed arrows in the figure:  protein synthesis and the initial folding to the intermediate state, and aggregation of the misfolded proteins.  Our focus will be on a single protein once it enters the intermediate state, and then transitions among the four states.  Similarly we ignore the aggregation process, occurring over much larger timescales than the transitions in the network.  We model the dynamics in the aftermath of a heat shock~\cite{eisen1998,richter2010}:  a sudden jump to some high temperature $T$, which then remains fixed as the system adapts.  The conditions favor misfolding over the native folding pathway.  In the absence of chaperones, state 2 (misfolded) would be most likely, and over longer timescales this would eventually result in a build-up of aggregates.  

\rev{To understand how this system can be controlled, let us summarize the various transitions in the network (Fig.~\ref{f1}A).  The protein can interconvert between states 3 and 1 with rates $k_m$ and $k_{-m}$.} A chaperone can bind to the misfolded protein at rate $k_c C(t)$, where $C(t)$ is the concentration of unoccupied chaperones and $k_c$ is the binding constant.  Once bound, the chaperone catalyzes the partial unfolding of the misfolded state to the intermediate state at rate $k_a$.  This conversion may involve several substeps and hydrolysis of multiple ATP molecules, but we simplify the process to a single reaction step hydrolyzing one ATP molecule, with some rate function $k_a$.  Though typically negligible compared to the forward rate $k_a$, the reverse rate $k_{-a}C(t)$, proportional to chaperone concentration, must be formally defined in order to have a thermodynamically complete description of the system.  Transitions from the intermediate to native state occur with rate $k_n$, and from the native to misfolded state with rate $k_u$.  \rev{The full details of the model, including parameter values estimated from experimental data on the chaperone GroEL assisting the folding of the substrate protein MDH~\cite{chakrabarti2017}, are described in Appendix~\ref{chapdet}.}

\rev{The model allows us to determine what kinds of external control are possible in the system.  We will consider two different scenarios:  the first is the typical case with one control knob, as seen in yeast, where chaperone concentration $C(t)$ can vary as part of the heat shock response, but ATP levels (and thus $k_a$) are fixed.  In principle changes in $C(t)$ affect two edges in the network, via the rates $k_c C(t)$ and $k_{-a} C(t)$.  However $k_{-a}$ is usually so small that changes in $C(t)$ make no noticeable difference to the current between states 2 and 3, which is dominated by the rate $k_a$.  For example in our parameter set $k_{-a}/k_a = 0.0952$ M$^{-1}$.  Since chaperone concentrations are usually on the $\mu$M scale or smaller, $k_{-a} C(t)$ at least seven orders of magnitude smaller than $k_a$.  If $k_a$ is fixed, we can treat the edge between states 2 and 3 as effectively non-controllable.  Hence in this scenario we have a network with a single controllable edge (edge 1 between states 1 and 2) and thus can choose a target subgraph as shown in Fig.~\ref{f1}B.  The target state in the subgraph is state 1, and state 2 is included as the non-target state to fulfill the local control condition.  We will dub this scenario one-state local control.  Given the danger of having a high probability of misfolded proteins, state 1 is a natural target for control.  One could imagine of course other biologically plausible targets, for example trying to control the probability of the protein being in its functional native state 4.  However given the available control knob $C(t)$, no target subgraph including state 4 would satisfy the local control condition, since there are no controllable edges involving state 4.  This highlights the usefulness of the condition to help us rationalize the influence of external factors on the system.

To illustrate how one-state local control works, we choose a target trajectory for $\rho_1(t)$ that is a rapid sigmoidal decrease in the misfolded probability over the course of a few minutes, shown as the dotted curve in Fig.~\ref{f1}D.  This would be a desirable heat shock response for the system.  Details of the target and the local control solution are described in Appendix~\ref{chapdet}.  Fig.~\ref{f1}C shows the necessary chaperone concentration protocol $\widetilde{C}(t)$ needed to drive state 1 along the target, derived from Eq.~\eqref{gex4}.  Note how $\widetilde{C}(t)$ closely resembles the qualitative features of actual chaperone expression in yeast experiments (Fig.~\ref{exp}A):  a large initial increase, a peak, and then a more gradual leveling off.  This kind of protocol is the typical solution if your target is a rapid suppression of the misfolded probability.  If we plug in the $\widetilde{C}(t)$ result and numerically solve the master equation for the system, we find $p_1(t)$ (solid blue curve in Fig.~\ref{f1}D) agreeing exactly with $\rho_1(t)$, as expected.  The remaining non-target state probabilities are compatible with Eq.~\eqref{gex1} for $\bm{\pi}(t) = (p_2(t), p_3(t), p_4(t))$.  As explained in Appendix~\ref{chapdet}, the second clause in the local control condition, that all components of $\bm{\pi}(t)$ must be non-negative during driving, actually gives us information about the kinds of target functions $\rho_1(t)$ that we can implement.  Virtually any $\rho_1(t)$ that is monotonically decreasing leads to non-negative $\bm{\pi}(t)$ solutions, regardless of how steep the decrease.  On the other hand, if one chose an increasing $\rho_1(t)$ one could violate the non-negative $\bm{\pi}(t)$ criterion.  This distinction agrees with our biological intuition:  changing chaperone concentration gives us fine-grained control in decreasing the misfolded probability, via the outgoing transition $k_c C(t)$ from state 1.  An arbitrary increasing $\rho_1(t)$ target would be biologically detrimental, and thus it is not surprising the system has not evolved the proper control knobs to achieve it.

In the second scenario we have two control knobs:  in addition to $C(t)$, we imagine ATP levels can change, as seen in {\it E. coli}, allowing $k_a(t)$ to be a time-varying function.  This gives two independently controllable edges (again ignoring the negligible role of $k_{-a}$), and allows us to choose a target subgraph as seen in Fig.~\ref{f1}E.  We fulfill the local control condition for two target states (1 and 2), and thus dub this scenario two-state local control.  Despite the additional control knob, we still cannot include state 4 as a target.  Having both $\rho_1(t)$ and $\rho_2(t)$ as targets allows one to control the total probability of observing a misfolded protein, both free (state 1) and bound to chaperone (state 2).  For example one can avoid the peak for state 2 seen in the one-state local control results of Fig.~\ref{f1}D.  Such transient accumulation of misfolded proteins bound to chaperones might not be ideal if the chaperones need to be turned over quickly to handle multiple different substrate proteins.  Fig.~\ref{f1}G shows the same $\rho_1(t)$ target as in the one-state control case, but with $\rho_2(t)$ chosen to be a small sigmoidal step, avoiding transient accumulation (red dotted curve).  The solutions for $\widetilde{C}(t)$ and $\widetilde{k}_a(t)$ are shown in Fig.~\ref{f1}F.  The chaperone concentration protocol is nearly identical, and $\tilde{k}_a(t)$ exhibits a transient peak, necessary to suppress the build-up of misfolded proteins on the chaperones.  Comparing with the {\it E. coli} experimental results in Fig.~\ref{exp}B, we again have qualitative similarities in both chaperone expression and ATP concentration behavior.  At least for our representative set of model parameters, the influence of the additional $k_a(t)$ control knob is not as large as that of $C(t)$.  This may in part explain why varying ATP levels is a fairly atypical heat shock response, while chaperone upregulation is universal.  However the two-state local control case does clearly illustrate the increasing precision of influence available with additional control knobs.
}

\section{Concluding remarks\label{con}}

Our theory of classical stochastic driving and its biological applications open up a variety of questions for future work.  \rev{We have focused here on discrete state Markov models, but taking the continuum limit of such models allows one to connect to diffusive dynamics described by Fokker-Planck equations.  In Appendix~\ref{fp} we show the simplest such connection, using the continuum limit of a 1D lattice to recover the Fokker-Planck CD driving theory of Refs.~\cite{li2017shortcuts,patra2017}.  However there are still open questions, like what a continuum version of local control would look like.}

The fact that there can exist many CD protocols for the same target trajectory, with distinct thermodynamic properties, means that one can search among these protocols for those that optimize certain quantities---like minimizing dissipated work under given physical constraints.  Optimal control of nonequilibrium and finite-time processes is an active research area~\cite{schmiedl2007optimal,sivak2012thermodynamic, Todd_near_optimal_protocols,Avishek_colloid}, with connections to techniques like Monge-Kantorovich transport theory~\cite{aurell2011optimal} and trajectory-observable biasing within the framework of large deviations \cite{Chetrite_Touchette_control,Avishek_Dom_rare}.  Situating \rev{our driving theory} within the broader context of these earlier optimal control approaches is an interesting topic for further study, both generically and in specific biological implementations in areas like ecology and evolution~\cite{nourmohammad2021optimal,lassig2020eco}.

Driving a system between long-lived states is also subject to universal bounds or ``speed limits''~\cite{kuznets2021dissipation,shiraishi2018speed,deffner2017quantum} that constrain the speed of driving in terms of dissipated work.  Does CD \rev{or non-CD global} driving saturate these bounds in certain circumstances?  If so, are there biological implications, for example cases where natural selection has pushed a control process close to the theoretical limit?  \rev{Finally, are there analogous bounds when we only specify local control targets?}

In summary, stochastic processes and their biological realizations are an ideal laboratory for investigating nonequilibrium control ideas.  The driving framework we have developed is a particularly useful starting point, because the control protocols can be expressed analytically in terms of easy-to-calculate graph properties of the underlying Markov model.  We can thus in principle explore a wide swath of \rev{control} solutions, and identify generic features of control in diverse biological systems sharing similar graph topologies.  The practicality of our formulation makes it well suited for deriving driving prescriptions in specific experimental contexts, like evolving cell populations~\cite{iram2021controlling} or the example systems in the current work.  Because the protocols involve accessible control knobs---\rev{like varying drug/protein concentrations}--- we believe near-term experimental validation is within reach.  \rev{Thus our approach may help with} implementing control of biological systems in the lab, and \rev{also} understanding how that control operates in nature. 

\acknowledgements{The authors would like to thank the stimulating environment provided by the Telluride Science Research Center, where this project was conceived. M.H. acknowledges support from the U.S. National Science Foundation (NSF) under Grant No. MCB-1651650. E.I. acknowledges support from the Labex CelTisPhyBio (ANR-11-LABX-0038, ANR-10-IDEX-0001-02).}

%\ef{It would also be interesting to add a discussion about potential future applications on deterministic rate equations in chemical system along the formalism of Ref.\cite{rao2016nonequilibrium}.}
\appendix

\section{\rev{Derivation of the local control solution}} \label{deriv}

\rev{Here we present a derivation of the local control solution, Eqs.~\eqref{gex1}-\eqref{gex4}.  The partitioning of the $\nabla$ and $G$ matrices described in Sec.~\ref{local} allows us to split Eq.~\eqref{f2} into two coupled equations,
\begin{eqnarray}
 \partial_t \bm{\rho}(t) &=& \nabla^{\rho c} \bm{\mathcal{\widetilde J}}{}^c(t)+\nabla^{\rho n} \bm{\mathcal{\widetilde J}}{}^n(t),\label{g6}\\
 \partial_t \bm{\pi}(t) &=& \nabla^{\pi c} \bm{\mathcal{\widetilde J}}{}^c(t)+\nabla^{\pi n} \bm{\mathcal{\widetilde J}}{}^n(t),\label{g7}
 \end{eqnarray}
 and similarly split Eq.~\eqref{m4ex} into
 \begin{eqnarray}
 \bm{\mathcal{\widetilde{J}}}{}^c(t) &=& \widetilde{G}^{c\rho}(t) \bm{\rho}(t) + \widetilde{G}^{c\pi}(t) \bm{\pi}(t),\label{g6b}\\
 \bm{\mathcal{\widetilde{J}}}{}^n(t) &=& \widetilde{G}^{n\rho}(t) \bm{\rho}(t) + \widetilde{G}^{n\pi}(t) \bm{\pi}(t).\label{g7b}
 \end{eqnarray}
 For generality we allow $\widetilde{G}^{n\rho}(t)$ and $\widetilde{G}^{n\pi}(t)$ to depend on time via some fixed (not externally modifiable) time-dependent functions for the non-controllable rates $\bm{\widetilde{k}}^{\pm n}(t)$.  However we will later specialize to the more typical case where $\bm{\widetilde{k}}^{\pm n}$ are time-independent.
 
 If the stretched inverse $[\nabla^{\rho c}]^{-1}_S$ exists, which can be determined from the graphical criterion described in Sec.~\ref{local}, then Eq.~\eqref{g6} can be inverted to find an expression for $\bm{\mathcal{\widetilde J}}{}^c(t)$,
 \begin{equation}
     \label{al1}
     \bm{\mathcal{\widetilde J}}{}^c(t) = [\nabla^{\rho c}]^{-1}_S \left(\partial_t\bm{\rho}(t)  -\nabla^{\rho n} \bm{\mathcal{\widetilde J}}{}^n(t)\right)+ C^c \bm{\Phi}^c(t).
 \end{equation}
Here $C^c$ is the matrix whose $\Delta_c$ columns are the basis vectors for the null space of $\nabla^{\rho c}$, and $\bm{\Phi}^c(t)$ is a $\Delta_c$-dimensional vector of arbitrary functions.  If we plug Eq.~\eqref{al1} into Eq.~\eqref{g7}, and substitute the right-hand side of Eq.~\eqref{g7b} for  $\bm{\mathcal{\widetilde{J}}}{}^n(t)$, we can rewrite Eq.~\eqref{g7} as
\begin{equation}
    \label{al2}
    \partial_t \bm{\pi}(t) = B \widetilde{G}^{n\pi}(t) \bm{\pi}(t) + \bm{a}(t),
\end{equation}
where
\begin{equation}\label{al3}
     \begin{split}
         B &= \nabla^{\pi n} - \nabla^{\pi c} [\nabla^{\rho c}]^{-1}_S \nabla^{\rho n},\\
         \bm{a}(t) &= \nabla^{\pi c}[\nabla^{\rho c}]^{-1}_S \partial_t \bm{\rho}(t) + B \widetilde{G}^{n\rho}(t) \bm{\rho}(t) + \nabla^{\pi c} C^c \bm{\Phi}^c(t).
     \end{split}
 \end{equation}
 Since the quantities that determine $B$, $\widetilde{G}^{n\pi}(t)$, and $\bm{a}(t)$ are known at the outset of the problem, we can always numerically solve the linear system of differential equations in Eq.~\eqref{al2} for $\bm{\pi}(t)$, given some initial condition $\bm{\pi}(0)$.  In the common scenario where the non-controllable rates are time-independent, and hence also the matrix $\widetilde{G}^{n\pi}$, Eq.~\eqref{al2} has an analytical solution, given by Eq.~\eqref{gex1}:
 \begin{equation}\label{al4}
 \bm{\pi}(t) = e^{t B \widetilde{G}^{n\pi}} \bm{\pi}(0) + \int_0^t dt^\prime\, e^{(t-t^\prime) B \widetilde{G}^{n\pi} } \bm{a}(t^\prime).
 \end{equation}

Once $\bm{\pi}(t)$ is known, the next step is to solve for $\bm{\mathcal{\widetilde J}}{}^c(t)$.  Substituting Eq.~\eqref{g7b} into Eq.~\eqref{al1}, we obtain the result of Eq.~\eqref{gex3}:
\begin{equation}
    \label{al5}
    \begin{split}
\bm{\mathcal{\widetilde J}}{}^c(t) =& [\nabla^{\rho c}]^{-1}_S \left(\partial_t\bm{\rho}(t)  - \nabla^{\rho n} \widetilde{G}^{n\rho} \bm{\rho}(t) - \nabla^{\rho n}\widetilde{G}^{n\pi}\bm{\pi}(t)\right)\\ &+ C^c \bm{\Phi}^c(t).
\end{split}
\end{equation}

Finally, we can relate the controllable edge currents $\bm{\mathcal{\widetilde J}}{}^c(t)$ to the corresponding edge rates $\bm{\widetilde{k}}^{\pm c}(t)$ via the analogue of Eq.~\eqref{sol2},
\begin{equation}
    \label{al6}
     {\bm{\mathcal{\widetilde J}}}{}^c(t) = M^{+c}(t) \bm{\widetilde{k}}^{+c}(t)  - M^{-c}(t) \bm{\widetilde{k}}^{-c}(t),
\end{equation}
where $M^{\pm c}(t) = \text{diag}({\nabla^{\pm \rho c}}^T \bm{\rho}(t) + {\nabla^{\pm \pi c}}^T \bm{\pi}(t))$.  The submatrices ${\nabla^{\pm \rho c}}$ and ${\nabla^{\pm \pi c}}$ are based on the same partition as shown in Fig.~\ref{control}A, except substituting $\nabla^\pm$ for $\nabla$.  Solving Eq.~\eqref{al6} for $\bm{\widetilde{k}}^{+ c}(t)$, we find Eq.~\eqref{gex4}:
\begin{equation}
    \label{al7}
    \bm{\widetilde{k}}^{+c}(t) = [M^{+c}(t)]^{-1} {\bm{\mathcal{\widetilde J}}}{}^c(t) + [M^{+c}(t)]^{-1} M^{-c}(t) \bm{\widetilde{k}}^{-c}(t).
\end{equation}

}

\section{Details of the repressor-copressor model calculations}\label{repdet}

The entire biochemical network of
Fig.~\ref{rep}A, including both clockwise and counterclockwise
transitions, was experimentally measured for PurR, and the parameters
are given by~\cite{xu1998kinetic}: $k_r = 0.0191$ nM$^{-1}$
min$^{-1}$, $k_c = 7.83 \times 10^{-4}$ nM$^{-1}$ min$^{-1}$, $k_x =
0.9$ nM$^{-1}$ min$^{-1}$, $k_{-r} = 1.68$ min$^{-1}$, $k_{-c} = 0.72$
min$^{-1}$, $k_{-x} = 0.072$ min$^{-1}$.  Note that $k_{-x} \ll
k_{-r}$ (the repressor-corepressor complex unbinds from the operator
more slowly than bare repressor) and $k_x \gg k_r$ (it binds more
easily), demonstrating the enhanced affinity of the complex to the
operator relative to the bare repressor.

While our system description focuses on the state
of the operator, the repressor and corepressor can also bind/unbind in
solution away from the operator~\cite{xu1998kinetic}, and in some
systems there are other molecules (like inducers) competing for
the repressor in solution.  \rev{In general then we will take the
solution concentrations $(R(t), C(t), X(t))$ to be some functions
determined by processes outside of the system, and explore how these three
control knobs can influence the state of the operator.

Following the graphical solution procedure of Sec.~\ref{sec:graphsol},
we start with the $N=3$, $E=3$ oriented current graph for the model, with currents oriented as shown in Fig.~\ref{rep}B,C.  The incidence matrix for the graph is
\begin{equation}\label{r0a}
\nabla = \begin{pmatrix}
-1 & 0 & -1\\
1 & -1 & 0\\
0 & 1  & 1\\
\end{pmatrix}.
\end{equation}
The matrix can be decomposed as $\nabla = \nabla^- - \nabla^+$ using Eqs.~\eqref{f2b}-\eqref{f3}, where
\begin{equation}\label{r0b}
\nabla^- = \begin{pmatrix}
0 & 0 & 0\\
1 & 0 & 0\\
0 & 1  & 1\\
\end{pmatrix}, \quad \nabla^+ = \begin{pmatrix}
1 & 0 & 1\\
0 & 1 & 0\\
0 & 0  & 0\\
\end{pmatrix}.
\end{equation}
The reduced incidence matrix $\widehat\nabla$ is given by the 
first two rows of Eq.~\eqref{r0a}.  Because $\Delta = E-N+1 = 1$, we have $\Delta + 1 = 2$ trees in a tree
basis.  Taking the tree with edge 1 missing as the reference (tree 1
in Fig.~\ref{rep}B), we choose the other tree in the basis to be the
one with edge 2 missing (tree 2 in Fig.~\ref{rep}C).  Using the
graphical algorithm, we can easily write down stretched inverse
reduced incidence matrices for these trees:
\begin{equation}\label{r3}
[\widehat\nabla^{(1)}]^{-1}_{S} = \begin{pmatrix}
  0 & 0\\
  0 & -1\\
  -1 & 0
\end{pmatrix}, \quad
[\widehat\nabla^{(2)}]^{-1}_{S} = \begin{pmatrix}
  0 & 1\\
  0 & 0\\
  -1 & -1
\end{pmatrix}.
\end{equation}
One can readily check that $\widehat\nabla
[\widehat\nabla^{(\gamma)}]^{-1}_{S}$ for $\gamma = 1,2$ is the
$2\times 2$ identity matrix.  There is a single fundamental cycle
vector for the graph, shown as a dashed line in Fig.~\ref{rep}C, given
by $\bm{c}^{(1)}=(1,1,-1)$.

Using Eq.~\eqref{sol1} we can write the currents ${\bm{\mathcal{\widetilde J}}}(t)$ to achieve a certain target $ \widehat{\bm{\rho}}(\lambda_t)$ as
\begin{equation}\label{r4}
  \begin{split}
    {\bm{\mathcal{\widetilde J}}}(t) &=[\widehat\nabla^{(1)}]^{-1}_S \partial_t \widehat{\bm{\rho}}(\lambda_t) + \Phi_1(t) \bm{c}^{(1)},
   \end{split}
\end{equation}
where $\Phi_1(t)$ is an arbitrary function.  The backward rates $\bm{\widetilde{k}}^-(t) = \bm{k}^- = (k_{-r}, k_{-c}, k_{-x})$ are fixed, and we can solve for the forward rates $\bm{\widetilde{k}}^+(t) = (k_r \widetilde{R}(t), k_c \widetilde{C}(t), k_x \widetilde{X}(t))$ using Eq.~\eqref{sol3}. The diagonal matrices $M^\pm(t) = \text{diag}({\nabla^\pm}^T \bm{\rho}(\lambda_t))$ in Eq.~\eqref{sol3} are given by
\begin{equation}
    \label{r4b}
    \begin{split}
    M^+(t) &= \begin{pmatrix} \rho_1(\lambda_t) & 0 & 0\\ 0 & \rho_2(\lambda_t) & 0 \\ 0 & 0 & \rho_1(\lambda_t)\end{pmatrix},\\
    M^-(t) &= \begin{pmatrix} \rho_2(\lambda_t) & 0 & 0\\ 0 & \rho_3(\lambda_t) & 0 \\ 0 & 0 & \rho_3(\lambda_t)\end{pmatrix}.
    \end{split}
\end{equation}
Substituting the expressions from Eqs.~\eqref{r3}-\eqref{r4b} into Eq.~\eqref{sol3}, we can solve for the concentration protocols $\widetilde R(t)$,
$\widetilde C(t)$, $\widetilde X(t)$ that determine the forward rates:
\begin{equation}\label{r6}
  \begin{split}
    \widetilde R(t) &= \frac{\Phi_1(t) + k_{-r}\rho_2(\lambda_t)}{k_r \rho_1(\lambda_t)},\\
    \widetilde C(t) &=\frac{\Phi_1(t) -\partial_t \rho_2(\lambda_t)  + k_{-c}\rho_3(\lambda_t)}{k_c \rho_2(\lambda_t)},\\
    \widetilde X(t) &=\frac{-\Phi_1(t) - \partial_t \rho_1(\lambda_t) + k_{-x}\rho_3(\lambda_t)}{k_x \rho_1(\lambda_t)}.
  \end{split}
\end{equation}
Different choices of $\Phi_1(t)$ correspond to different control protocols
that drive the system through the same trajectory $\bm{\rho}(\lambda_t)$.  For example $\Phi_1(t) = 0$ gives the protocol associated with tree 1 (Fig.~\ref{rep}B), and $\Phi_1(t) = \partial_t \rho_2(\lambda_t)$ gives the protocol associated with tree 2 (Fig.~\ref{rep}C).  The
one additional constraint is that only $\Phi_1(t)$ functions that lead to
non-negative concentrations in Eq.~\eqref{r6} at all $t$ during driving are physically allowable.

To illustrate a family of control protocols, we need to choose a specific target trajectory $\bm{\rho}(\lambda_t)$.  Based on the discussion in Sec.~\ref{nonstat}, Eq.~\eqref{r6} describes the control protocol regardless of whether the target trajectory is an instantaneous stationary one or not.  However to be concrete, we will choose an instantaneous stationary trajectory, making our control protocols CD.  To mimic a rapid switch in gene expression from on to off, we will select $\bm{\rho}(\lambda_t)$ to be the stationary distribution associated with a set of concentration functions that serve as control parameters in the original system, $\lambda_t = (C(t), R(t), X(t))$.  We choose $C(t)$ to sharply increase in a sigmoidal fashion, with $R(t)$ kept at a constant level and $X(t)$ in detailed balance with $C(t)$ and $R(t)$:
\begin{equation}\label{r7}
\begin{split}
    R(t) &= R_0, \quad C(t) = C_0 + (C_f - C_0) \frac{e^{k(t-t_0)}}{1+e^{k(t-t_0)}},\\
    X(t) &= \frac{k_r R(t) k_c C(t) k_{-x}}{k_{-r} k_{-c} k_x},
\end{split}
\end{equation}
where $R_0 = 20$ nM, $C_0 = 0.2$ $\mu$M, $C_f = 20$ $\mu$M, $k = 3$ min$^{-1}$, $t_0 = 5$ min, and the remaining parameters are described above.    Eq.~\eqref{r7} determines the forward rates $\bm{k}^+(t)$ that enter into the transition matrix $\Omega(\lambda_t)$, and hence allows us to use Eq.~\eqref{2} to solve for the instantaneous stationary state $\bm{\rho}(\lambda_t)$.  The components of this stationary target distribution are shown in Fig.~\ref{repd}A.  They represent a transition from a system dominated by state 1 at the beginning of the protocol to one dominated by state 3 at the end (the gene turning mostly off).

The local detailed balance and optimal control protocols shown in Fig.~\ref{repd} can be found numerically in a straightforward way.  For the detailed balance case, we need to satisfy the condition $\widetilde{\chi}^\text{cyc}(t) = C^T \bm{\widetilde{\chi}}(t) = 0$, where $\widetilde{\chi}^\text{cyc}(t)$ is a scalar because $\Delta =1$.  Plugging in the expressions from Eq.~\eqref{r6}, the $\widetilde{\chi}^\text{cyc}(t)=0$ condition becomes a nonlinear equation for $\Phi_1(t)$.  We can then use numerical root finding to determine the $\Phi_1(t)$ at each $t$ that satisfies the condition, thus defining the local detailed balance protocol.  In a similar way, the optimal protocol (minimizing entropy production) should satisfy Eq.~\eqref{t2}.  When we plug Eq.~\eqref{r6} into Eq.~\eqref{t2}, we get another nonlinear equation for $\Phi_1(t)$, which can be numerically solved at each $t$.  We verified that the resulting $\Phi_1(t)$ is exactly the same as what we would get by direct numerical minimization of $\dot{S}^\text{tot}(t)$ in Eq.~\eqref{cd1}.  We also checked that the solution for $\Phi_1(t)$ always gives non-negative rates $\bm{\widetilde{k}}^+(t)$, or equivalently non-negative concentrations $\widetilde{R}(t)$, $\widetilde{C}(t)$, and $\widetilde{X}(t)$.

}

\section{Details of the chaperone model calculations}\label{chapdet}

\subsection{\rev{Transition rates and model parameters}}

\rev{The transition rates of the chaperone model in Fig.~\ref{f1}A satisfy certain local detailed balance relationships.  The rates $k_m$ and $k_{-m}$, describing interconversion between states 3 and 1, obey}
\begin{equation}\label{c0}
\frac{k_{-m}}{k_m} = e^{-\beta \epsilon_{m}},
\end{equation}
where $\epsilon_m>0$ is the free energy difference between the intermediate and misfolded states.  \rev{The rates $k_{-n}$ and $k_{-u}$ are related to their counterparts $k_n$ and $k_{u}$ through}
\begin{equation}\label{c2}
\frac{k_{-n}}{k_n} = e^{-\beta \epsilon_{n}}, \qquad \frac{k_{-u}}{k_u} = e^{-\beta \epsilon_{u}}.
\end{equation}
Here $\epsilon_{n}$ and $\epsilon_{u}$ are the free energy differences between the intermediate and native, and between the native and misfolded states respectively.  Since going from states \rev{1 $\to$ 4 $\to$ 3} should yield the same cumulative free energy difference as going directly from \rev{1 $\to$ 3}, we know that $\epsilon_{m} = \epsilon_{u} + \epsilon_{n}$.

Since a full traversal of the left loop clockwise (states \rev{3 $\to$ 1 $\to$ 2 $\to$ 3}) involves hydrolysis of an ATP molecule, the product of clockwise/counterclockwise rate ratios over the entire cycle is related to the chemical potential difference $\Delta \mu$ of ATP hydrolysis:
\begin{equation}\label{c1}
\frac{k_m k_c C(t) k_a}{k_{-m} k_{-c} k_{-a} C(t)} = \frac{k_m k_c k_a}{k_{-m} k_{-c} k_{-a}}  = e^{\beta \Delta \mu}.
\end{equation}

We base the parameter values in our model on those associated with the chaperone GroEL assisting the folding of the substrate protein MDH, estimated from fitting to experimental data~\cite{chakrabarti2017}:  $k_m=0.37$ min$^{-1}$, $k_n = 0.366$ min$^{-1}$, $k_u = 0.025$ min$^{-1}$, $k_{-u} = 7.78 \times 10^{-3}$ min$^{-1}$, $k_c = 1.7 \times 10^6$ M$^{-1}$min$^{-1}$, $k_a = 4$ min$^{-1}$.  In cases where only upper or lower bounds on the parameters could be determined, we used the values at the bound.  Using Eq.~\eqref{c2} and the values of $k_u$ and $k_{-u}$ yield an estimate of $\epsilon_u = 1.17$ $k_B T$.  We do not know the precise value of $\epsilon_m$ from the experimental fitting, but we assume a typical value of $\epsilon_m = 3$ $k_B T$, which then gives $\epsilon_n = \epsilon_m - \epsilon_u = 1.83$ $k_BT$.  Similarly, we set $k_{-c} = 0.1$ min$^{-1}$ as the unbinding rate of the chaperone, a typical scale assuming strong binding affinity between the chaperone and substrate.  The remaining unknown parameters can now be determined using Eqs.~\eqref{c0}-\eqref{c1} (setting the ATP hydrolysis potential difference $\Delta \mu = 22$ $k_BT$~\cite{milo2015cell}):  $k_{-m} = 0.0184$ min$^{-1}$, $k_{-n} = 0.0585$ min$^{-1}$, \rev{$k_{-a} = 0.381$ M$^{-1}$min$^{-1}$.  As mentioned in the Sec.~\ref{chapmodel}, given typical chaperone concentrations $C(t) \sim  O(1\:\mu\text{M})$ or smaller, we get $k_{-a} C(t) \ll k_a$, and hence we can neglect the effect of the $k_{-a}$ transition.}

\subsection{\rev{One-state local control}}

\rev{Let us first consider the one-state local control solution corresponding to the target subgraph of Fig.~\ref{f1}B.  Here the number of target states $N_T =1$, and hence we want $p_1(t) = \rho_1(t)$ for some chosen target trajectory $\rho_1(t)$, while the non-target states are given by $\bm{\pi}(t) = (p_2(t), p_3(t), p_4(t))$.  The number of controllable edges is $E_c = 1$, with only edge 1 in Fig.~\ref{f1}B amenable to external control (assuming fixed $k_a$ and negligible $k_{-a}$).  The goal is to solve for the controllable edge rate vector $\bm{\widetilde{k}}^{+c}(t) = (k_c \widetilde{C}(t))$ from Eq.~\eqref{gex4}, using the method outlined in Sec.~\ref{local}.  The various quantities needed to construct the solution are as follows.  The submatrices of $\nabla$ and $\widetilde{G}(t)$ are given by
\begin{equation}\label{loc1}
\begin{split}
&\nabla^{\rho c} = \begin{pmatrix} -1 \end{pmatrix},\quad \nabla^{\rho n} = \begin{pmatrix} 0 & 1 & 0 & 1 \end{pmatrix},\\
&\nabla^{\pi c} = \begin{pmatrix} 1\\ 0\\ 0 \end{pmatrix},\quad \nabla^{\pi n} = \begin{pmatrix} -1 & 0 & 0 & 0\\
1 & -1 & -1 & 0\\
0 & 0 & 1 & -1
\end{pmatrix},
\end{split}
\end{equation}
and
\begin{equation}\label{loc2}
\begin{split}
&\widetilde{G}^{c \rho}(t) = \begin{pmatrix} k_c \widetilde{C}(t) \end{pmatrix},\quad \widetilde{G}^{c \pi} = \begin{pmatrix} -k_{-c} & 0 & 0 \end{pmatrix},\\
&\widetilde{G}^{n \rho} = \begin{pmatrix} 0\\ -k_{-m}\\ 0 \\ -k_{-u} \end{pmatrix},\quad \widetilde{G}^{n\pi} = \begin{pmatrix} k_a & 0 & 0\\
0 & k_m & 0\\
0 & k_n & -k_{-n}\\
0 & 0 & k_u
\end{pmatrix},
\end{split}
\end{equation}
where we have set $k_{-a} \approx 0$ in $\widetilde{G}^{n\pi}$ because its effect is negligible.  Because the target subgraph associated with $\nabla^{\rho c}$ is tree-like, the stretched inverse $[\nabla^{\rho c}]^{-1}_S$ is just the ordinary inverse: $[\nabla^{\rho c}]^{-1}_S = (-1)$.  Given these submatrices, we can calculate the matrix $B \widetilde{G}^{n\pi}$ and vector $\bm{a}(t)$ needed to evaluate the expression for $\bm{\pi}(t)$ in Eq.~\eqref{gex1}:
\begin{equation}\label{loc3}
\begin{split}
&B \widetilde{G}^{n\pi} = \begin{pmatrix} 
-k_a & k_m & k_u\\
k_a & -k_m-k_n & k_{-n}\\
0 & k_n & -k_u-k_{-n}
\end{pmatrix},\\
&\bm{a}(t) = \begin{pmatrix}
-\partial_t \rho_1(t) - (k_{-u}+k_{-m}) \rho_1(t)\\
k_{-m}\rho_1(t)\\
k_{-u}\rho_1(t)
\end{pmatrix}.
\end{split}
\end{equation}
Note that there is no dependence on arbitrary functions $\bm{\Phi}^c(t)$ in $\bm{a}(t)$, since $\nabla^{\rho c}$ has no null space ($\Delta_c = E_c - N_T = 0$, and hence $C^c$ does not exist).  The integral in Eq.~\eqref{gex1} can then be carried out numerically to find $\bm{\pi}(t)$.  Knowing $\bm{\pi}(t)$ allows us to evaluate the currents at the controllable edges $\bm{\mathcal{\widetilde J}}{}^c(t)$ from Eq.~\eqref{gex3}.  Finally we plug these currents into Eq.~\eqref{gex4} to find $\bm{\widetilde{k}}^{+c}(t)$ and hence our control protocol $\widetilde{C}(t)$.  The matrices that appear in this equation are $M^{+c}(t) = ( \rho_1(t) )$, $M^{-c}(t) = (\pi_1(t))$, and the reverse rate vector is $\bm{\widetilde{k}}^{-c} = (k_{-c})$, which cannot be externally varied.

To get the control results shown in Fig.~\ref{f1}C,D we chose a sigmoidal decreasing target function for $\rho_1(t)$ of the form
\begin{equation}\label{loc4}
\rho_1(t) = \alpha_1 \tanh(\kappa(t-t_0)) + \beta_1,
\end{equation}
where $\alpha_1 = -0.276$, $\beta_1 = 0.368$, $\kappa=1$ min$^{-1}$, and $t_0 = 5$ min.  We numerically integrated Eq.~\eqref{gex1} for a range of $t$ from 0 to 20 min, choosing $\bm{\pi}(0) = (0.003,0.054,0.299)$ initial conditions compatible with the value of $\rho_1(0) = 0.644$.  Note that in the absence of any control protocol, our starting probability distribution at $t=0$ is the stationary distribution corresponding to a low ($10^{-2}$ $\mu$M) concentration of chaperones.  Of course once the local control protocol begins, the system does not follow a stationary distribution trajectory, since this is not a global CD solution.

To understand when Eq.~\eqref{gex1} for $\bm{\pi}(t)$ yields invalid solutions (components $\pi_i(t) < 0$ for some $t$ during driving), we can look at the structure of the system of differential equations, Eq.~\eqref{al2}, whose solution is Eq.~\eqref{gex1}.  To prevent $\pi_i(t)$ from becoming negative, the derivative $\partial_t \pi_i(t)$ evaluated at $\pi_i(t) = 0$ must be non-negative.  Using Eq.~\eqref{al2}, and plugging in the expressions from Eq.~\eqref{loc3} we can thus write three conditions:
\begin{equation}\label{loc5}
    \begin{split}
    \left.\partial_t \pi_1(t)\right|_{\pi_1(t)=0} &= -\partial_t \rho_1(t) - (k_{-u} + k_{-m})\rho_1(t)\\
    &\qquad + k_m \pi_2(t) + k_u \pi_3(t) \ge 0,\\
    \left.\partial_t \pi_2(t)\right|_{\pi_2(t)=0} &= k_{-m} \rho_1(t) + k_a \pi_1(t) + k_{-n} \pi_3(t) \ge 0,\\
    \left.\partial_t \pi_3(t)\right|_{\pi_3(t)=0} &= k_{-u} \rho_1(t) + k_n \pi_2(t) \ge 0.
    \end{split}
\end{equation}
The second and third conditions are automatically fulfilled, since all the terms on the right-hand side are non-negative by construction.  Only the first condition can sometimes be violated.  For our parameter set $k_{-u}+k_{-m} = 0.026$ min$^{-1}$ is small relative to $k_m = 0.37$ min$^{-1}$, so the only term likely to cause trouble is $-\partial_t \rho_1(t)$.  However if the target $\rho_1(t)$ is monotonically decreasing, we get a positive contribution to the right-hand side and can generally fulfill the condition.  Thus local control will be possible for a wide range of biologically plausible target functions where the goal is suppressing the misfolded probability.

\subsection{\rev{Two-state local control}}

To find the two-state local control solution corresponding to the target subgraph of Fig.~\ref{f1}E, we proceed analogously to the one-state solution described above.  We now have $N_T =2$ and thus two target functions $\rho_1(t)$ and $\rho_2(t)$ for states 1 and 2.  The non-target states are $\pi(t) = (p_3(t), p_4(t))$.  The number of controllable edges $E_c = 2$, and we seek solutions for the controllable edge rate vector $\bm{\widetilde{k}}^{+c}(t) = (k_c \widetilde{C}(t), \widetilde{k}_a(t))$ 
from Eq.~\eqref{gex4}.  The submatrices of $\nabla$ and $\widetilde{G}(t)$ are
\begin{equation}\label{loc1b}
\begin{split}
&\nabla^{\rho c} = \begin{pmatrix} -1 & 0\\ 1 & -1\end{pmatrix},\quad \nabla^{\rho n} = \begin{pmatrix} 1 & 0 & 1\\ 0 & 0 & 0 \end{pmatrix},\\
&\nabla^{\pi c} = \begin{pmatrix} 0 & 1\\ 0 & 0\end{pmatrix},\quad \nabla^{\pi n} = \begin{pmatrix} -1 & -1 & 0\\
0 & 1 & -1
\end{pmatrix},
\end{split}
\end{equation}
and
\begin{equation}\label{loc2b}
\begin{split}
&\widetilde{G}^{c \rho}(t) = \begin{pmatrix} k_c \widetilde{C}(t) & -k_{-c} \\ 0 & \widetilde{k}_a(t)\end{pmatrix},\quad \widetilde{G}^{c \pi} = \begin{pmatrix} 0 & 0 \\ 0 & 0 \end{pmatrix},\\
&\widetilde{G}^{n \rho} = \begin{pmatrix} -k_{-m} & 0\\ 0 & 0 \\ -k_{-u} &0 \end{pmatrix},\quad \widetilde{G}^{n\pi} = \begin{pmatrix} 
k_m & 0\\
k_n & -k_{-n}\\
0 & k_u
\end{pmatrix},
\end{split}
\end{equation}
where we have again set $k_{-a} \approx 0$.  As before, the target subgraph associated with $\nabla^{\rho c}$ is tree-like, so the stretched inverse $[\nabla^{\rho c}]^{-1}_S$ is just the ordinary inverse:
\begin{equation}\label{loc2c}
[\nabla^{\rho c}]^{-1}_S = \begin{pmatrix} -1 & 0 \\ -1 & -1\end{pmatrix}.
\end{equation}
We also have $\Delta_c = E_c - N_T =0$, so $C^c$ does not exist.  The resulting expressions for $B \widetilde{G}^{n\pi}$ and vector $\bm{a}(t)$ that enter into Eq.~\eqref{gex1} are:
\begin{equation}\label{loc3b}
\begin{split}
&B \widetilde{G}^{n\pi} = \begin{pmatrix} 
-k_n & k_u + k_{-n}\\
k_n & -k_u - k_{-n}
\end{pmatrix},\\
&\bm{a}(t) = \begin{pmatrix}
-\partial_t \rho_1(t) -\partial_t \rho_2 - k_{-u} \rho_1(t)\\
k_{-u}\rho_1(t)
\end{pmatrix}.
\end{split}
\end{equation}
The matrices $M^{\pm c}(t)$ that appear in Eq.~\eqref{gex4} are:
\begin{equation}
    \label{loc3c}
    M^{+c}(t) = \begin{pmatrix}
    \rho_1(t) & 0\\
    0 & \rho_2(t)
    \end{pmatrix},\quad
    M^{-c}(t) = \begin{pmatrix}
    \rho_2(t) & 0\\
    0 & \pi_1(t)
    \end{pmatrix}.
\end{equation}
To plot the control protocol results of Fig.~\ref{f1}F,G, we choose the same $\rho_1(t)$ target function from Eq.~\eqref{loc4}.  For $\rho_2(t)$ we also choose a sigmoidal target:
\begin{equation}\label{loc4b}
\rho_2(t) = \alpha_2 \tanh(\kappa(t-t_0)) + \beta_2,
\end{equation}
where $\alpha_2 = 0.008$, $\beta_2 = 0.011$, $\kappa=1$ min$^{-1}$, and $t_0 = 5$ min.

Just as in the one-state local control case, we can derive conditions for having valid $\bm{\pi}(t)$ solutions from the structure of Eq.~\eqref{al2}:
\begin{equation}\label{loc5b}
    \begin{split}
    \left.\partial_t \pi_1(t)\right|_{\pi_1(t)=0} &= -\partial_t \rho_1(t) -\partial_t \rho_2(t) - k_{-u} \rho_1(t) \\
    &\qquad + (k_u + k_{-n}) \pi_2(t) \ge 0,\\
    \left.\partial_t \pi_2(t)\right|_{\pi_2(t)=0} &= k_{-u} \rho_1(t) + k_n \pi_1(t) \ge 0.
    \end{split}
\end{equation}
The second condition is always fulfilled, while the first one depends on the target trajectory.  Since $k_{-u} \ll k_u$, the main concern is again the time derivatives of the target functions, $\partial_t \rho_1(t) + \partial_t \rho_2(t)$.  However if the total misfolded target probability $\rho_1(t) + \rho_2(t)$ (free and bound to chaperone) is monotonically decreasing, two-state local control solutions generally exist.
}

\section{CD driving in continuous systems}\label{fp}

Let us consider discrete-state Markov models on lattice graphs (also known as grid graphs). In these cases the states can be visualized as points on some $d$-dimensional lattice, with transitions occurring between neighboring lattice points.  If we imagine the states as actual positions in a $d$-dimensional space, and allow the lattice spacing to become infinitesimal as the number of states $N \to \infty$, then the behavior of such models should approach continuum diffusive dynamics described by Fokker-Planck equations.  Thus, taking appropriate limits, we should be able to use our formalism to derive \rev{control} solutions for Fokker-Planck systems.  \rev{Here we describe how to do this for a $d=1$ lattice in the global CD control case, rederiving the Fokker-Planck CD driving results of Refs.~\cite{li2017shortcuts,patra2017}.}  Beyond this validation, we demonstrate how CD driving works for systems exhibiting position-dependent diffusivity, not considered in Refs.~\cite{li2017shortcuts,patra2017}.

To connect our \rev{formalism} to Fokker-Planck dynamics, let us first describe a one-dimensional Fokker-Planck equation for the time evolution of a probability density $p(x,t)$,
\begin{equation}\label{fp11}
\partial_t p(x,t) = -\frac{\partial}{\partial x} \left[A(x)p(x,t) \right] + \frac{\partial^2}{\partial x^2} \left[D(x)p(x,t)\right],
\end{equation}
where $x$ is our position variable, $A(x,t)$ is the drift function, and $D(x)$ is the position-dependent local diffusivity.  Though $D(x)$ is often taken to be a constant, $D(x) = D$, \rev{we} allow it to be position-dependent for generality.  We focus on the case where the drift $A(x,t) = -D(x) \partial_x E(x,\lambda_t)$, and hence arises from forces due to a potential energy $E(x,\lambda_t)$ that may be dependent on time-varying control parameters $\lambda_t$.  Eq.~\eqref{fp11} can then be rewritten as
\begin{equation}
\label{fp11ex}
\begin{split}    
\partial_t p(x,t) &=-\frac{\partial}{\partial x} \left[-D(x) \rho(x,\lambda_t) \frac{\partial}{\partial x} \frac{p(x,t)}{\rho(x,\lambda_t)} \right],\\
&\equiv -\frac{\partial}{\partial x} J(x,t),
\end{split}
\end{equation}
where
\begin{equation}\label{fp12}
\rho(x,\lambda_t) = \frac{e^{-\beta E(x,\lambda_t)}}{Z(\lambda_t)}.
\end{equation}
From the structure of Eq.~\eqref{fp11ex} it is clear that $\rho(x,\lambda_t)$ is the instantaneous stationary distribution that makes the right-hand side vanish.  We assume the energy function $E(x,\lambda_t) \to \infty$ as $x \to x_L$ and $x \to x_R$, defining a domain of $x$ of width $\Delta x = x_R - x_L$.  Thus the partition function $Z(\lambda_t) = \int_{x_L}^{x_R} dx\,\exp(-\beta E(x,\lambda_t))$ is well-defined.  An infinite domain would correspond to the special case where $\Delta x \to \infty$.  The second line of Eq.~\eqref{fp11ex} defines a probability current density $J(x,t)$, in terms of which the Fokker-Planck equation takes the form of a continuity equation for probability.

\begin{figure}
    \centering
    \includegraphics[width=\columnwidth]{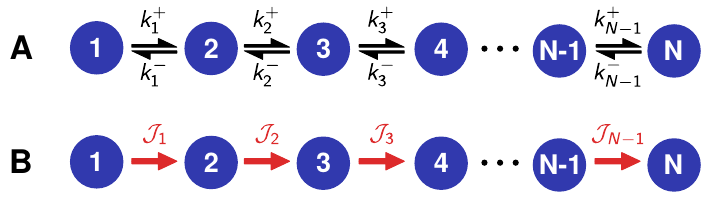}
    \caption{A) $N$-state Markov model on a one-dimensional lattice graph, with $E = N-1$ edges.  Black arrows correspond to transitions between neighboring states, \rev{$k_i^+(\lambda_t)$ and $k_i^-(\lambda_t)$}, $i=1,\ldots,N-1$, which depend on the control protocol $\lambda_t$.  B) Oriented stationary currents $\mathcal{J}_\alpha(t)$, $\alpha = 1,\ldots,N-1$.  These currents form the only spanning tree for the graph.}
    \label{chain}
\end{figure}

To apply our \rev{global CD control approach} for discrete Markov systems, let us construct a one-dimensional lattice graph Markov model with $N$ states, shown in Fig.~\ref{chain}A, that approximates the Fokker-Planck equation as $N \to \infty$.  State $i$ corresponds to position $x_i=x_L +ia$, where $a=\Delta x/N$ is the lattice spacing, which becomes infinitesimal for large $N$.  In this limit the probability $p_i(t)$ of being in state $i$ is related to the probability density $p(x,t)$ through $a^{-1} p_i(t) \to p(x_i,t)$.

In the discrete model the nonzero transition matrix elements correspond to the forward (right) arrows, $\Omega_{i+1,i}(\lambda_t) = \rev{k^+_i(\lambda_t)}$, and the backward (left) arrows, $\Omega_{i,i+1}(\lambda_t) = \rev{k^-_i(\lambda_t)}$, for $i=1,\ldots,N-1$.  We choose the following forms for the transition rates~\cite{bicout1998}:
\begin{equation}\label{fp1}
\begin{split}
\rev{k^+_i(\lambda_t)}&=\frac{D_i}{a^2}e^{-\frac{1}{2}\beta(E_{i+1}(\lambda_t)-E_i(\lambda_t))},\\
\rev{k^-_i(\lambda_t)}&=\frac{D_i}{a^2}e^{\frac{1}{2}\beta(E_{i+1}(\lambda_t)-E_i(\lambda_t))}.
\end{split}
\end{equation}
Here $D_i \equiv D(x_i)$ and $E_i(\lambda_t) \equiv E(x_i,\lambda_t)$ 
are the discrete versions of the local diffusivity and potential energy. The ratio of the forward and backward transitions satisfies the local detailed balance relationship
\begin{equation}\label{fp2}
\begin{split}
\rev{\frac{k_i^+(\lambda_t)}{k_i^-(\lambda_t)}}&=e^{-\beta(E_{i+1}(\lambda_t)-E_i(\lambda_t))}.
\end{split}
\end{equation}
As a result, the instantaneous stationary distribution for this system assumes a form analogous to Eq.~\eqref{fp12},
\begin{equation}\label{fp3}
\begin{split}
\rho_i(\lambda_t) &=\frac{e^{-\beta E_i(\lambda_t)}}{{\cal Z}(\lambda_t)},
\end{split}
\end{equation}
where ${\cal Z}(\lambda_t) = \sum_{i=1}^N \exp(-\beta E_i(\lambda_t))$.  To check whether the transition rates of Eq.~\eqref{fp1} give the correct Fokker-Planck equation in the continuum limit, we note that the master equation for the discrete system can be written as:
\begin{equation}
    \label{fpe1}
\partial_t p_i(t) = \sum_j \Omega_{ij}(\lambda_t) p_j(t) = -J_{i+1}(t) + J_i(t),
\end{equation}
where the current from state $i$ to $i+1$ is given by
\begin{equation}
    \label{fpe2}
J_{i}(t) = \rev{k_i^+(\lambda_t)} p_i(t) - \rev{k_i^-(\lambda_t)} p_{i+1}(t).
\end{equation}
Eq.~\eqref{fpe1} is the discrete analogue of the second line in Eq.~\eqref{fp11ex}, with the conversion $J_i(t) \to J(x_i,t)$, $a^{-1} p_i(t) \to p(x_i,t)$.  Plugging Eq.~\eqref{fp1} into Eq.~\eqref{fpe2}, we can rewrite the current $J_i(t)$ as
\begin{equation}
    \label{fpe3}
J_{i}(t) = -\frac{1}{a} D_i \sqrt{\rho_{i+1}(\lambda_t) \rho_i(\lambda_t)}\frac{1}{a}\left[ \frac{p_{i+1}(t)}{\rho_{i+1}(\lambda_t)} -\frac{p_i(t)}{\rho_i(\lambda_t)} \right].
\end{equation}
Eq.~\eqref{fpe3} goes to the correct limit in the continuum case, becoming the current density in the square brackets in Eq.~\eqref{fp11ex}.  To see this, note that
\begin{equation}\label{fpe3b}
\begin{split}
&a^{-1}\sqrt{\rho_{i+1}(\lambda_t) \rho_i(\lambda_t)}\\
&\qquad \to \left[\rho(x_i+a,\lambda_t)\rho(x_i,\lambda_t) \right]^{1/2}\\
&\qquad \approx\left[\left(\rho(x_i,\lambda_t) + a \partial_x\rho(x_i,\lambda_t)\right)\rho(x_i,\lambda_t)\right]^{1/2}\\
&\qquad =\rho(x_i,\lambda_t) + {\cal O}(a),
\end{split}
\end{equation}
where ${\cal O}(a)$ denotes corrections of order $a$.  Thus Eq.~\eqref{fp1} is a valid discretization of the Fokker-Planck system.  It is not unique (other choices are possible, as shown in the SI) but any valid discretization should lead to the same CD results in the continuum limit.

With the discretization validated, we can now proceed to applying the general solution procedure.  The oriented current graph ($N$ states, $E = N-1$ edges) is tree-like, so the graph itself is the only spanning tree. Using the graphical algorithm we can write down the $(N-1) \times (N-1)$ dimensional stretched inverse reduced incidence matrix for this tree,
\begin{equation}\label{fp5}
[\widehat\nabla^{(1)}]^{-1}_{S} = \begin{pmatrix}
-1 & 0 & 0 & 0 &\cdots & 0\\
-1 & -1 & 0 & 0 & \cdots & 0\\
-1 & -1 & -1 & 0 & \cdots & 0\\
-1 & -1 & -1 & -1 & \cdots & 0\\
\vdots & \vdots & \vdots & \vdots & \ddots & \vdots\\
-1 & -1 & -1 & -1 & \cdots & -1
\end{pmatrix}.
\end{equation}
Because the graph is tree-like, the stretched inverse is also the ordinary inverse of the reduced incidence matrix, $[\widehat\nabla^{(1)}]^{-1}_{S} = \widehat\nabla^{-1}$.  From Eqs.~\eqref{fp1} and \eqref{fp3} we can deduce that the stationary currents have zero magnitude:
\begin{equation}\label{fp4}
\begin{split}
\mathcal{J}_i(t) &=\rev{k_i^+(\lambda_t)}\rho_i(\lambda_t) - \rev{k_i^-(\lambda_t)}\rho_{i+1}(\lambda_t) = 0.
\end{split}
\end{equation}
Hence we know that $\bm{\mathcal{\widetilde{J}}}(t)= \delta\bm{\mathcal{J}}(t)$.  Moreover, since there are no cycles in the graph, Eq.~\eqref{gs3} gives us the full CD current solution:
\begin{equation}\label{fp6}
\bm{\mathcal{\widetilde{J}}}(t) = [\widehat\nabla^{(1)}]^{-1}_S \partial_t \widehat{\bm{\rho}}(\lambda_t).
\end{equation}

Let us assume CD rates $\widetilde{k}^+_i(t)$ and $\widetilde{k}^-_i(t)$ of a form analogous to Eq.~\eqref{fp1},
\begin{equation}\label{fp7}
\begin{split}
\rev{\widetilde{k}_i^+(t)}&=\frac{\widetilde{D}_i(t)}{a^2}e^{-\frac{1}{2}\beta(\widetilde{E}_{i+1}(t)-\widetilde{E}_i(t))},\\
\rev{\widetilde{k}_i^-(t)}&=\frac{\widetilde{D}_i(t)}{a^2}e^{\frac{1}{2}\beta(\widetilde{E}_{i+1}(t)-\widetilde{E}_i(t))},
\end{split}
\end{equation}
where $\widetilde{D}_i(t)$ represents a modified, potentially time-dependent, local diffusivity which we allow for generality, and $\widetilde{E}_{i}(t)$ is the energy associated with state $i$ in the CD protocol.  In many cases it may not be possible to control the local diffusivity via external parameters, and hence it remains unchanged, $\widetilde{D}_i(t) = D_i$.  However as will be seen from the structure of the CD solution described below, we have in principle the freedom to choose $\widetilde{D}_i(t)$ to be any non-negative function.  The energy perturbation at each site due to the CD protocol is $U_i(t) = \widetilde{E}_i(t) - E_i(\lambda_t)$.  To solve for these CD perturbations $U_i(t)$, the first step is to rewrite Eq.~\eqref{fp6} using Eq.~\eqref{fp5} and the expression for $\bm{\mathcal{\widetilde{J}}}(t)$ in terms of the CD transition rates:
\begin{equation}\label{fp8}
\rev{\widetilde{k}_i^+(t)}\rho_i(\lambda_t) - \rev{\widetilde{k}_i^-(t)}\rho_{i+1}(\lambda_t) = -\sum_{j=1}^i \partial_t \widehat{\rho}_j(\lambda_t).
\end{equation}
After plugging in Eq.~\eqref{fp7} for the CD rates, and Eq.~\eqref{fp3} for the stationary distribution, Eq.~\eqref{fp8} can be written as:
\begin{equation}\label{fp9}
\begin{split}
&-2a^{-2}\widetilde{D}_i(t) \sqrt{\widehat{\rho}_i(\lambda_t)\widehat{\rho}_{i+1}(\lambda_t)} \sinh\left[\frac{\beta(U_{i+1}(t) - U_{i}(t))}{2}\right]\\
&\qquad\qquad =-\sum_{j=1}^i \partial_t \widehat{\rho}_j(\lambda_t).
\end{split}
\end{equation}
We can invert this to find a recursion relation for the $U_i(t)$,
\begin{equation}\label{fp10}
U_{i+1}(t) -U_i(t) = \frac{2}{\beta}\sinh^{-1}\left[\frac{a^{2} \sum_{j=1}^i \partial_t \widehat{\rho}_j(\lambda_t)}{2 \widetilde{D}_i(t) \sqrt{\widehat{\rho}_i(\lambda_t)\widehat{\rho}_{i+1}(\lambda_t)}}\right].
\end{equation}
Given an arbitrary choice of function $U_1(t)$ (which corresponds to the freedom of redefining the zero level for energies), we can use consecutive applications of Eq.~\eqref{fp10} to solve for $U_i(t)$, $i=2,\ldots,N$.

The final step is to transform the CD results back to the continuum, where the CD energies can be expressed as $\widetilde{E}(x,t) = E(x,\lambda_t) + U(x,t)$.  The perturbations $U(x,t)$ can be found from the continuum analogue of Eq.~\eqref{fp10},
\begin{equation}\label{fp13}
\frac{\partial U(x,t)}{\partial x} = \frac{1}{\beta \widetilde{D}(x,t) \rho(x,t)} \int_{x_L}^x dx^\prime \partial_t \rho(x^\prime,\lambda_t).
\end{equation}
To derive this we have expanded in small $a$ and used the fact that $\sinh^{-1}(\epsilon) \approx \epsilon$ to lowest order in $\epsilon$.  In the continuum limit $\sum_{j=1}^i a \,\partial_t \widehat{\rho}_j(\lambda_t) \to \int_{x_L}^x dx^\prime \partial_t \rho(x^\prime,\lambda_t)$ and $a^{-1} \sqrt{\widehat{\rho}_i(\lambda_t) \widehat{\rho}_{i+1}(\lambda_t)} \to \rho(x,\lambda_t)$, to leading order.  This follows from the same argument as Eq.~\eqref{fpe3b}, setting $x=x_i$.

From Eq.~\eqref{fp13} we can directly solve for $U(x,t)$,
\begin{equation}\label{fp14}
\begin{split}
&U(x,t) = U_0(t)\\
&\qquad+ \int_{x_0}^x dx^\prime \frac{1}{\beta \widetilde{D}(x^\prime,t) \rho(x^\prime,\lambda_t)} \int_{x_L}^{x^\prime} dx^{\prime\prime} \partial_t \rho(x^{\prime\prime},\lambda_t),
\end{split}
\end{equation}
where $x_0$ is an arbitrary reference position and $U_0(t)$ is an arbitrary energy offset function (which does not affect the driving).

In practice, a particular CD protocol means simultaneously implementing the diffusivity $\widetilde{D}(x,t)$ and perturbing the energy landscape by $U(x,t)$.  As mentioned earlier, in many experimental scenarios control of diffusivity will not be possible, so the only available CD protocols will involve keeping the diffusivity equal to the value in the original system, $\widetilde{D}(x,t) = D(x)$.  One special case of this is a position-independent diffusivity $D(x) = D$ that is not varied during the CD protocol.  This was solved by Li {\it et al.}~\cite{li2017shortcuts} and Patra \& Jarzynski~\cite{patra2017} using alternative approaches, and their expressions for the CD perturbation are equivalent to our Eq.~\eqref{fp13} with the substitution $\widetilde{D}(x,t) = D$.  

From the perspective of thermodynamic costs, Eq.~\eqref{cd1} for our discrete-state system takes the form
\begin{equation}\label{fp15}
\begin{split}
&\dot{S}^\text{tot}(t)\\
&\qquad= k_B \sum_{i=1}^{N-1} \mathcal{\widetilde{J}}_i(t) \ln \frac{\rev{\widetilde{k}_i^+(t)} \rho_i(\lambda_t)}{\rev{\widetilde{k}_i^-(t)} \rho_{i+1}(\lambda_t)}\\
&\qquad= - \frac{1}{T} \sum_{i=1}^{N-1} \mathcal{\widetilde{J}}_i(t) (U_{i+1}(t) - U_i(t))\\
&\qquad= 2 k_B \sum_{i=1}^{N-1}  \mathcal{\widetilde{J}}_i(t) \sinh^{-1} \left[\frac{a^2 \mathcal{\widetilde{J}}_i(t)}{2 \widetilde{D}_i(t) \sqrt{\widehat{\rho}_i(\lambda_t)\widehat{\rho}_{i+1}(\lambda_t)}} \right],
\end{split}
\end{equation}
where we have used the CD rates from Eq.~\eqref{fp7} and $\mathcal{\widetilde{J}}_i(t) = -\sum_{j=1}^i \partial_t \widehat{\rho}_j(\lambda_t)$ from Eqs.~\eqref{fp5}-\eqref{fp6}.  The functional form for $\dot{S}^\text{tot}(t)$ is always non-negative, since $y \sinh^{-1}(c y) \ge 0$ for any $y$ when $c\ge0$.  In the limit of adiabatically slow driving, $\partial_t \widehat{\rho}_j(\lambda_t) \to 0$, we see that $\mathcal{\widetilde{J}}_i(t)\to 0$ and hence the entropy production rate $\dot{S}^\text{tot}(t) \to 0$.  As noted in Sec.~\ref{costs}, under the (unlikely) scenario that one can control the local diffusivity $\widetilde{D}_i(t)$ and make it large during the CD protocol, then $\dot{S}^\text{tot}(t)$ can be made small even for fast driving.

In the continuum limit, Eq.~\eqref{fp15} becomes
\begin{equation}\label{fp16}
    \dot{S}^\text{tot}(t)=k_B \int_{x_L}^{x_R} dx\, \frac{\mathcal{\widetilde{J}}^2(x,t)}{\widetilde{D}(x,t) \rho(x,t)},
\end{equation}
where $\mathcal{\widetilde{J}}_i(t) \to \mathcal{\widetilde{J}}(x,t)$ is the continuum CD current.  This expression has the same form as the standard Fokker-Planck result for $\dot{S}^\text{tot}(t)$~\cite{seifert2012stochastic}, with the CD current and CD local diffusivity substituted for the original ones.

\vspace{0.5in}

%\bibliography{counterdiabatic_refs.bib}


\end{document}